\documentclass[prx,twocolumn,english,longbibliography,superscriptaddress,floatfix,nobalancelastpage,showkeys]{revtex4-2}
\pdfoutput=1 % keep for arxiv, remove for physical review submission

\usepackage{stylesheet}
\usepackage{graphicx,subcaption}
\usepackage{dsfont}

\begin{document}

\notoc

\title{Quantum spherical codes}
% \thanks{A footnote to the article title}

\author{Shubham~P.~Jain}
\affiliation{\QUICS}

\author{Joseph~T.~Iosue}
\affiliation{\QUICS}
\affiliation{\JQI}

\author{Alexander~Barg}
\affiliation{\ECE}

\author{Victor~V.~Albert}
%\email{vva@umd.edu}
\affiliation{\QUICS}

\date{\today}

%\keywords{quantum error correction, bosonic codes, spherical codes, spherical designs}

\begin{abstract}
We introduce a framework for constructing quantum codes defined on spheres by recasting such codes as quantum analogues of the classical spherical codes.
We apply this framework to bosonic coding, obtaining multimode extensions of the cat codes
that can outperform previous constructions while requiring a similar type of overhead.
Our polytope-based cat codes consist of sets of points with large separation that at the same time form averaging sets known as spherical designs.
We also recast concatenations of CSS codes with cat codes as quantum spherical codes, \vva*{revealing a new way to autonomously protect against dephasing noise.}
\end{abstract}
%Our ``quantum spherical codes'' can be realized in any quantum system whose phase space contains a real or complex sphere.

\maketitle

% \tableofcontents

%The only currently known method to implement fault-tolerant quantum computation with arbitrary accuracy is with a family of quantum error-correcting codes defined on a many-body systems \cite{}. The simplest unit of such a system is the qubit, but

\eczoo[Bosonic (a.k.a.\@ oscillator) codes]{oscillators} %\cite{joshi_quantum_2021,girvin_introduction_2023,albert_bosonic_2022} 
offer alternative qubit blueprints %useful and feasible %logical
that are compatible with continuous-variable (CV) quantum platforms \cite{sun_tracking_2014,ofek_extending_2016,leghtas_confining_2015,touzard_coherent_2018,grimm_stabilization_2020,campagne-ibarcq_quantum_2020,fluhmann_encoding_2019,de_neeve_error_2022,sivak_real-time_2022,dahan_creation_2022}
%\cite{girvin_introduction_2021,albert_bosonic_2022} 
%in hybrid coding schemes 
and that can reduce overhead by offering an extra layer of protection
\cite{cohen_dissipation-induced_2014,fukui_analog_2017,fukui_high-threshold_2018,guillaud_repetition_2019,vuillot_quantum_2019,noh_fault-tolerant_2020,chamberland_building_2022,larsen_fault-tolerant_2021,guillaud_error_2021,noh_low-overhead_2022,zhang_concatenation_2022,regent_high-performance_2022,stafford_biased_2022,lieu_candidate_2022,Gouzien_Computing_2023}.
Qubits defined on a few bosonic modes or more exotic spaces \cite{albert_robust_2020} are likely to prove useful as control of quantum systems improves, but the field remains relatively unexplored \cite{byron_bay_quantum_computing_workshop_gkp_2021,gottesman_opportunities_2022}
in part because structures and intuition from qubit-based coding theory need not apply.

We develop a framework that yields generalizations of a class of bosonic codes called the \eczoo[\textit{cat codes}]{cat}
\cite{cochrane_macroscopically_1999,leghtas_hardware-efficient_2013} and unifies such codes with several others.
Our key observation is that all such codes are particular instances of quantum versions of \eczoo[\textit{spherical codes}]{spherical} \cite{conway_sphere_1999,ericson_codes_2001}, a family well known in classical coding theory.
%Spherical codes are useful for communication because of a famous observation by Shannon \cite{} (see also \cite{}) that the space of energy-constrained electromagnetic signals can be mapped onto a high-dimensional sphere.
%We develop several infinite families of codes as well as few-mode instances based on polytopes that are more powerful than current alternatives.
% underlying Hilbert-space structure of such codes is different from the digital Pauli-based structure of qubit codes
% Such codes are quite different
% The underlying Hilbert-space structure makes their analysis quite different from
% A class of codes known as the cat codes \cite{} were the first to achieve
%Unify cat codes, pair-cat codes, diatomic molecular codes, and certain large-spin codes under one formalism by making contact and generalizing a well-known class of classical codes --- the spherical codes \cite{}.
%In this short paper,
We overview the framework and demonstrate its utility with several new multimode cat codes.
%Focus on coherent-state QSCs, demonstrating the formalism
%We leave the general group-theoretic framework for future work.
A rigorous study of general features
%of the framework
is left to a companion follow-up work. 

\parg{General codes on the sphere}
% A codeword of  qubit code is constructed by taking a quantum superposition of a set of bit strings.
% By analogy, a quantum spherical codeword is a quantum superposition of points taken from a spherical code, i.e., a set, or \textit{constellation}, of points on the unit sphere.
% We consider uniform superpositions here, leaving more general studies to future work.
% A quantum spherical code, or QSC, is then defined using a collection of $K$ \textit{logical constellations} that make up the \textit{code constellation}, $\CC=\bigcup_{k=1}^{K}\CC_{k}$.

Codewords of \eczoo[qubit codes]{qubits_into_qubits} are quantum superpositions of bit-strings. 
By analogy, we start with a spherical code, which is a set, or constellation, of points on the unit sphere.
To construct a \textit{quantum spherical code}, or \textit{QSC}, we take a collection $\{\CC_k\}_{k=0}^{K-1}$ of \textit{logical constellations}, each of which gives rise to a codeword of the QSC obtained by taking a quantum superposition of all points $\mathbf{x}\in \CC_k$. 
We consider uniform superpositions here, leaving more general codes to future work. 
Taken together, the logical constellations yield the \textit{code constellation}, $\CC=\bigcup_{k=0}^{K-1}\CC_{k}$.

% A spherical code is a finite set, or \textit{constellation}, of points on the $n$-dimensional unit sphere.
% %\redout{with each codeword corresponding to a point of the constellation.}
% By analogy to \eczoo[qubit codes]{qubits_into_qubits}, whose
% codewords are quantum superpositions of bit-strings, we take the $k$th codeword
% of a \textit{quantum spherical code}, or \textit{QSC}, to be a quantum superposition
% of points in its corresponding spherical \textit{logical constellation} $\CC_k$.
% We consider uniform superpositions here, leaving more general studies
% to future work.
% The full QSC is a set of $K$ logical constellations
% that make up the \textit{code constellation}, $\CC=\bigcup_{k=1}^{K}\CC_{k}$.

In the electromagnetic setting, spherical codes protect classical information against signal fluctuations during transmission, which correspond to small shifts acting on
points in the constellation.
A code's ability to protect against such errors can be quantified by the minimum (squared) Euclidean distance $\de$ between any pair of distinct points.
%in the constellation.
%QSCs store quantum information in the $K$-dimensional subspace spanned by states that are superpositions of points of logical constellations.
QSCs 
%also provide protection against such errors, 
naturally inherit $\de$ as a figure of merit for protecting against such ``bit-flip'' noise.

Since QSCs store quantum information, they also suffer from ``phase'' noise, which comes from, e.g., fiber attenuation.
Such noise can be expressed in terms of ``potential-energy'' functions on the sphere whose evaluation can be used to distinguish logical constellations (cf. [\citealp{albert_robust_2020}, Sec.~VI.B; \citealp{pedernales_decoherence-free_2020}]).
If the average of a function over points
in a constellation ${\cal C}_{k}$ depends on $k$, then the function's underlying physical process causes an undetectable ``phase'' error.
%A general notion of distance quantifying protection from such errors depends on the physical system embedding the QSC.
%, but since
%functions on the sphere can be expanded in spherical harmonics, a
%reasonable notion of $\dd$ can count the maximum degree of harmonics
%that yields a $k$-independent average over logical constellations.

\begin{figure}[!t]
\centering{}
\includegraphics[width=1.0\columnwidth]{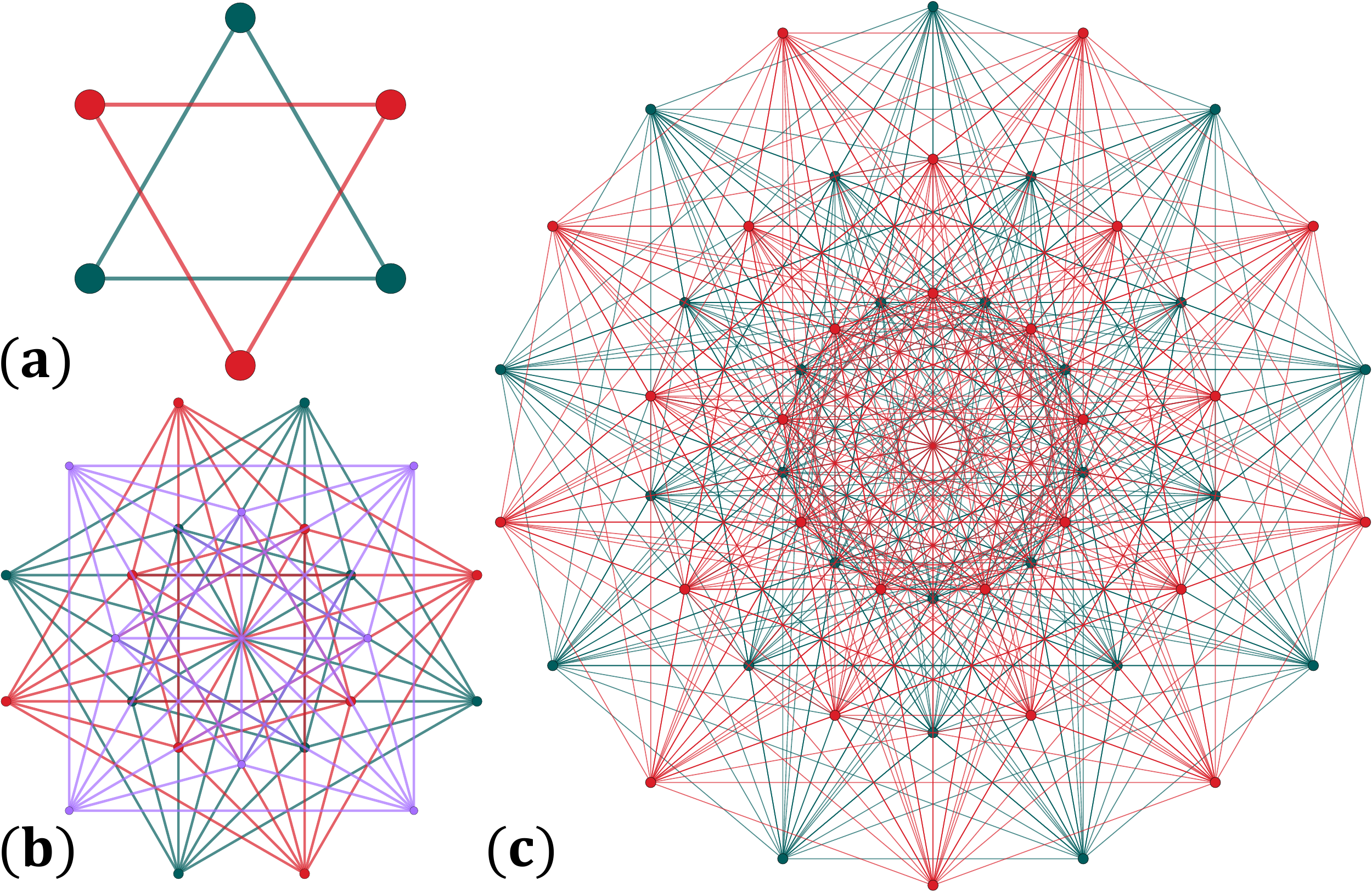}
\caption{
\label{fig:qsc}
Quantum spherical codewords are quantum superpositions of constellations on a sphere.
Logical constellations can form the vertices of a polytope and unite to form a code polytope compound. Projections of polytope compounds are shown for the {$\mathbf{(a)}$} cat, $\mathbf{(b)}$ Möbius-Kantor, and $\mathbf{(c)}$ Hessian quantum spherical codes, with logical constellation points colored either green, red, or purple.
}
\end{figure}

An $(n,K,\de)$ spherical code contains $K$ points
on the $n$-dimensional unit sphere such that the squared Euclidean distance
between any two points is at least $\de$.
An $\sph{n,K,\de,\dots}$ QSC
is a $K$-dimensional
subspace of a quantum system's vector space whose
%canonical
%/computational basis
states
are labeled by points on an $n$-dimensional
(real or complex)
unit sphere, and whose protection
against rotations is quantified by $\de$.
Protection against ``phase'' noise is designated by the proxy ``$\dots$'' because the notion of a ``phase-flip'' distance depends on the physical system embedding the QSC. 
%and whose protection against ``phase'' noise is quantified by realization-dependent parameter(s) 

In principle, the above framework applies to any quantum state space parameterized by points on a sphere.
%, and there several actively studied quantum systems to which this framework applies.
CV systems \cite{cerf_quantum_2007,serafini_quantum_2017} admit several such spaces, and there exist examples of QSCs expressed using ordinary \cite{cochrane_macroscopically_1999,leghtas_hardware-efficient_2013,denys_2t-qutrit_2022}, squeezed \cite{schlegel_quantum_2022,xu_autonomous_2022,hillmann_quantum_2022}, and pair-coherent states \cite{albert_pair-cat_2019}.
Collective atomic systems described by spin-coherent states as well as rotational state spaces of diatomic molecules also admit QSCs, namely, various \eczoo[large-spin codes]{single_spin} \cite{gross_encoding_2021} and diatomic molecular codes \eczoo{diatomic_molecular} \cite[Sec. VI]{albert_robust_2020}, respectively.
%Orientations of a planar rotor (heteronuclear diatomic molecule) are labeled by points in the real (one-) two-dimensional sphere, so molecular codes \cite{} and other formulations of robust rotational systems \cite{} are compatible with the QSC notion.
%Codes described by coherent \cite{}, spin-coherent \cite{}, or pair-coherent \cite{} states can all be thought of as QSCs.
We focus on coherent-state QSCs because
such codes naturally generalize the cat codes, and error-correction procedures for these new multimode cat codes require a similar type of overhead as what has already been realized \cite{sun_tracking_2014,ofek_extending_2016,leghtas_confining_2015,touzard_coherent_2018,grimm_stabilization_2020}.
% those that have already --- a candidate for a logical qubit architecture \cite{} with a well-developed technology.
We note that the discussion below can be modified to apply to other manifestations of QSCs.

\parg{Coherent-state formalism}
A single-mode \textit{coherent state} is a quantum representation of a standing wave of a fixed-frequency signal.
%(a.k.a. single-mode) signal.
An $n$-mode coherent state $|\av\ket$ is parameterized by a complex $n$-dimensional point $\av$. The point's norm $\Vert\av\Vert^2$ corresponds to the state's energy, and points of all states with a fixed energy $\nb$ form a complex $n$-sphere, $\Omega_n=\{\av\in\bbC^{n}\,,\,\Vert\av\Vert^2=\nb\}$.

Coherent-state QSCs consist of \spj*{disjoint} logical constellations $\CC_k$ of $|\CC_k|$ points picked from the $n$-sphere and superimposed to form logical codewords,
\begin{equation}\label{eq:coherent_states}
    |\CC_{k}\ket\sim\frac{1}{\sqrt{|{\cal C}_{k}|}}\sum_{\av\in\CC_{k}}|\sqrt{\nb}\av\ket~,
\end{equation}
where we restrict logical constellations to lie on the \textit{unit} $n$-sphere and \spj*{delegate the overall scaling of the sphere's radius to $\nb$.}
An example to keep in mind is the
%($n=1$)
%cat code --- the
%four-component
%BPSK-like \cite{} classical code consisting of two logical constellations $\CC_0=\{+1\}$ and $\CC_1=\{-1\}$.
%QPSK-like \cite{}
four-component cat code defined by $\CC_0=\{(1),(-1)\}$ and $\CC_1=\{(\i),(-\i)\}=\i\CC_0$.

The normalization in Eq. \eqref{eq:coherent_states} is \vva*{only} valid asymptotically as $\nb\to\infty$ because
coherent states are not quite orthogonal due to the uncertainty principle,
%Overlap between two participating coherent states decays exponentially with the separation of their corresponding points,
\begin{equation}\label{eq:codeword}
    |\bra\av|\bv\ket|^{2}=\exp\left(-\Vert\av-\bv\Vert^{2}\right)\leq\exp\left(-\nb d_{E}\right)~.
\end{equation}
The above ``quantum corrections'' for two coherent states of a code are suppressed exponentially with the energy $\nb$ and the minimum distance between two points in the code's constellation $\CC=\bigcup_{k} \CC_k$,
\begin{equation}
    \de = \min_{\av\neq\bv\in\CC} \Vert\av-\bv\Vert^2~.
\end{equation}
%These corrections imply that we cannot fully resolve constellation points.
% with the scale of resolution set by $\de$.
%protection against rotation errors is approximate, with intrinsic misidentification of codewords suppressed exponentially with increasing energy.
%will discuss this issue later on,
%ignore this issue for now and continue to express results asymptotically.
Since $\de$ sets the scale of resolution of the constellation points, we refer to it as the \textit{resolution} from now on.
% Since such corrections are quickly overcome as $\nb\to\infty$, our subsequent claims as well as the aforementioned Eq. (\ref{eq:coherent_states}) are valid only in this limit, requiring the use of asymptotic notation ``$\sim$''.

Coherent states are subjected to two essentially different types of distortion --- angular dephasing due to fluctuations in a mode's frequency and changes in the mode's excitations \cite[Sec.~II.A]{Turchette_decoherence_2000}.
These induce ``bit'' and ``phase'' noise on QSCs, respectively.
The corresponding relevant noise operators are passive linear-optical transformations and products of modal ladder operators $\{a_j,a_j^\dagger\}_{j=1}^n$, whose commutator is $[a_j,a_\ell^{\dg}]=\delta_{j\ell}$. 
%\joe{One of these should be daggered right?}
Products of
%passive linear-optical
transformations and ladder operators can be used to express any physical noise channel \cite[Eq. (39)]{grimsmo_quantum_2020}.

%Passive linear-optical
Transformations on $n$ modes are parameterized by the unitary group $\grp{U}(n)$ \cite[Sec. 5.1.2]{serafini_quantum_2017}.
A transformation $U_{\R}$
% \begin{equation}
%     U_{\R} = \exp(\half\i \sum)
% \end{equation}
corresponding to the $n$-dimensional unitary matrix $\R$ rotates a coherent state $|\av\ket$ into $|\R\av\ket$.
%according to $U_{\R} |\av\ket = |\R\av\ket$.
%, where ``$.$'' represents matrix-vector multiplication.
%Any transformation for which
If the rotation satisfies $\Vert \R\av - \av\Vert^2 < \de$, the transformation is detectable in the $\nb\to\infty$ limit. Codes with larger resolution protect against larger sets of transformations.
%The resolution yields a sufficient condition for a transformation to be detectable in the $\nb\to\infty$ limit, namely, $\Vert \R\av - \av\Vert^2 < \de$.
%Intrinsic memory errors due to imperfect resolution at finite $\nb$ are suppressed exponentially with $\nb\de$.

A general ladder error,
\begin{equation}\label{eq:ladder}
    L_{\pp,\qq}(\aa^{\dg},\aa)=\prod_{j=1}^{n}a_{j}^{\dg p_{j}}a_{j}^{q_{j}}~,
\end{equation}
is a monomial in the operators $(a_1,a_2,\cdots,a_n)=\aa$ and their adjoints. It
is parameterized by non-negative integer vectors $\pp=(p_1,p_2,\cdots,p_n)$ and $\qq=(q _1,q_2,\cdots,q_n)$ quantifying how many energy carriers (e.g., photons or phonons) are gained and lost in each mode, respectively.
%The one-norms $|\pp|,|\qq|$ are the degrees of $L$ in the variables $\{\a_j^\star\}$ and $\{\a_j\}$, respectively, signifying how many photons were gained and lost in all modes.

Lowering operators $a_j$ are ``diagonal'' in the coherent-state basis, satisfying $a_j|\av\ket=\a_j|\av\ket$, where $\a_j$ is the $j$th component of $\av$.
%Similarly, by taking the adjoint, we have $\bra\av|a^\dg_j=\a_j^\star \bra \av|$.
This ``diagonality'' relation and its adjoint imply that the expectation value of a ladder error over the $k$th codeword \eqref{eq:coherent_states} reduces to the average of the operator's corresponding monomial over $\CC_k$,
\begin{equation}\label{eq:ladder_kl}
    \bra{\CC}_{k}|L_{\pp,\qq}(\aa^{\dg},\aa)|{\CC}_{k}\ket\sim\frac{\nb^{|\pp+\qq|/2}}{|\CC_k|}\sum_{\av\in{\CC}_{k}}L_{\pp,\qq}(\av^{\star},\av)~,
\end{equation}
where the one-norm $|\pp+\qq|$ is the degree of $L_{\pp,\qq}(\av^{\star},\av)$.
%A QSC can detect a particular ladder error if the operator's expectation value cannot be used to distinguish the codewords.
A ladder error can be detected whenever the above average is \textit{independent} of $k$ \cite{knill_theory_1997}.

\parg{Polytope QSCs}
We have found numerous QSCs whose constellations form vertices of real \cite{coxeter_regular_1973} or complex \cite{shephard_regular_1952,coxeter_regular_1991} \eczoo[polytopes]{polytope}.
Polytope vertices are both sufficiently well-separated and uniform, providing protection against both types of noise.
%, i.e., regular polytopes whose vertices are complex
%(as opposed to real)
Code and polytope tables for the two cases are in Appxs.~\ref{appx:poly_real} and \ref{appx:poly_complex}, respectively.
% The codes are presented in Tables \ref{tab:polytope_qsc} and \ref{tab:polytope_qsc_complex}, and the relevant polytopes are collected in Tables \ref{tab:polytope} and \ref{tab:polytope_complex}, respectively.

We characterize ladder-error protection of polytope QSCs with three ``distances'': $d_\downarrow$, $t_\downarrow$, and $d_\updownarrow$.
%of total detectable losses (plus one, by coding theory convention) as well as the number $t_\downarrow$ of total correctable losses (plus one).
The first is the number of detectable losses (plus one), signifying that any pure-loss ladder error $L_{\pp=\boldsymbol{0},\qq}$ with $|\qq|<d_{\downarrow}$ is detectable.
Similarly, $t_{\downarrow}$ is the number of \textit{correctable} losses (plus one), signifying that any ladder error %$L_{\pp,\qq}$
with $|\pp|,|\qq|<t_{\downarrow}$ is detectable.
The
%third parameter --- the
\textit{degree distance} $d_{\updownarrow}$ signifies that the code detects ladder errors with degree $|\pp+\qq|<d_\updownarrow$.
% which is the maximum total degree $|\pp+\qq|$ of a detectable ladder error (plus one).
These three parameters satisfy
% the relations
\begin{equation}
    \lfloor(d_{\updownarrow}+1)/2\rfloor\leq t_{\downarrow}\leq d_{\updownarrow}\leq d_{\downarrow}
\end{equation}
and can vary quite significantly.

Our notation for an $n$-mode polytope QSC with $K$ logical codewords is $\sph{n,K,\de,d_{\updownarrow}}$ or, more generally,
$\sph{n,K,\de,\bra t_{\downarrow},d_{\updownarrow},d_{\downarrow}\ket}$.
%Below, we highlight the strengths of a few constructions relative to the cat codes.
%, denoting an , resolution $\de$, and aforementioned protection against ladder errors.
%, can tolerate rotations displacing points by no farther than $\sqrt{\de}$, can correct (detect) up to $t_{\downarrow}-1$ ($d_{\downarrow}-1$) losses, and can detect any ladder operator whose total degree is $d_{\updownarrow}-1$ or less.
The four-component cat code
%, $\CC_0=\{(\pm 1)\}$ and $\CC_1=\i \CC_0$,
is a $\sph{1,2,2.0,\bra 2,2,2 \ket}$ QSC, detecting $d_{\downarrow}-1=1$ loss error while sporting the relatively high resolution of $2.0$.
Since it can detect one gain simultaneously with one loss, this code also corrects $t_{\downarrow}-1=1$ loss error.
%\joe{Confusing myself here. If distance $d_\downarrow$ is only 2, how does it correct one loss error?}

Each logical constellation of the four-component cat code is a line segment, and the code constellation
%$\CC_0\cup\CC_1$
forms the vertices of a square.
More generally, logical constellations of the $2p$-component $\sph{1,2,4\sin^{2}\frac{\pi}{2p},\bra p,p,p\ket}$ cat code are two $p$-gons whose vertices interleave for maximal resolution.
There is a tradeoff between loss protection and resolution, with the latter of order $O(1/p^2)$ for a large number $p-1$ of correctable losses.
Utilizing higher dimensions, we pick other complex polytopes that maintain the same resolution while offering increased loss protection over the cat codes.

A simple code straddling the $p=2,3$ cat codes in terms of performance is the $\sph{2,2,1.5,\bra 2,3,3\ket}$ \textit{simplex code}, \begin{equation}\label{eq:simplex}
    \CC_0=\left\{\textstyle{\frac{1}{\sqrt{2}}}(\o^{\mu},\o^{2\mu})\,|\,\mu\in\grp{\bbZ_{5}}\right\}=-\CC_1~,
\end{equation}
where $\o = e^{i\frac{2\pi}{5}}$.
This code admits a lower resolution than the $p=2$ cat code, but detects one more loss in any of the two modes.
Equivalently, it admits a higher resolution than the $p=3$ cat code's resolution of unity, but corrects one fewer loss.
Simplices exist in any dimension, yielding the infinite $\sph{n,2,2-1/n,3}$ QSC family
that approaches the resolution of the $p=2$ cat code with increasing $n$
while detecting one more loss in any mode.

The \textit{Möbius-Kantor} $\sph{2,3,1.0,\bra 3,4,4\ket}$ code maintains the resolution of the $p=3$ cat code while adding one more logical state and detecting one more loss.
Each of its three logical constellations form the 8 vertices of a Möbius-Kantor polygon ($3\{3\}3$ in Coxeter notation; see Appx.~\ref{appx:poly_complex}), and such polygons combine to form the 24 vertices of a $3\{4\}3$ polygon.
This code corrects one more loss than the $\grp{2T}$\eczoo[-qutrit]{2t_qutrit} \cite{denys_2t-qutrit_2022}, a $\sph{2,3,1.0,\bra 2,4,4\ket}$ QSC whose logical constellations each make up the 8 vertices of a complex octagon $2\{4\}4$.
These two codes differ despite the fact that both code constellations map to the vertices of the \textit{same} real 4D polytope via the mapping $(x+iy,z+iw)\to(x,y,z,w)$, demonstrating subtleties in using real polytopes to define complex QSCs.

Logical constellations of the powerful $\sph{3,2,1.0,\bra 4,5,9 \ket}$ \textit{Hessian code} consist of the 27 vertices of a Hessian polytope,
\begin{equation}\label{eq:hessian}
  \CC_{0}=\left\{\textstyle{\frac{1}{\sqrt{2}}}(\eta^{\mu},-\eta^{\nu},0)\cup\text{perms.}\,|\,\mu,\nu\in\grp{\bbZ_{3}}\right\}=-\CC_1~,
\end{equation}
%and $\CC_1 = -\CC_0$,
where $\eta = e^{i\frac{2\pi}{3}}$, and ``$\text{perms.}$'' is shorthand for the two cyclic permutations of the vector to the left for each $\mu,\nu$.
This code corrects as many losses as the $p=4$ cat code, but has the resolution of the $p=3$ cat code. Moreover, it can detect up to 8 losses, a feature available only to the $p\geq9$ cat codes.

There is a $\sph{2,2,2-\sqrt{2},\bra5,6,12\ket}$ code that maintains the same resolution as the $p=4$ cat code, but corrects one more and detects 8 more losses.
Its logical constellations each form the 24 vertices of a $4\{3\}4$ polygon, combining into a 48-vertex $2\{6\}4$ polygon.

An overachieving cousin of the above code is the $\sph{4,2,2-\sqrt{2},\bra6,8,12\ket}$ \textit{Witting code}, consisting of two Witting polytopes with 240 vertices each.
This code corrects as many losses as a $p=6$ cat code, has the resolution of a $p=4$ cat code, and detects up to 11 losses.
It is the first member of the infinite $\sph{2^r,2,2-\sqrt{2},8}$ family of codes that are based on orbits of the real \eczoo[Clifford group]{sidelnikov} \cite{sidelnikov_finite_1997-1,sidelnikov_finite_1997,sidelnikov_spherical_1999,nebe_invariants_2001}.

%The Witting code is the first member of the infinite family of $\sph{2^r,2,2-\sqrt{2},8}$ \textit{Clifford group-orbit codes} for $r\geq 2$.
% Their name comes from the fact that each $\CC_k$
% %is a \textit{group orbit} of the real Clifford group, i.e., all of $\CC_k$
% can be created by applying unitary rotations making up the real Clifford group on a fiducial point \eczoo{sidelnikov} \cite{sidelnikov_finite_1997-1,sidelnikov_finite_1997,sidelnikov_spherical_1999,nebe_invariants_2000}.

A lower bound on $d_{\updownarrow}$ for Clifford, simplex, or other QSCs can be obtained whenever their logical constellations form designs \cite{delsarte_spherical_1977}.
% of strength $d_{\updownarrow}-1$.
A constellation $\CC_k$ is a \textit{complex spherical design} \cite{roy_complex_2014,mohammadpour_complex_2023} of strength $\tau$ if averages of monomials $L_{\pp,\qq}$ of total degree $|\pp+\qq|\leq\tau$ over $\CC_k$ \eqref{eq:coherent_states} are equal to those over the entire unit $n$-sphere,
\begin{equation}
  \frac{1}{|\CC_{k}|}\sum_{\av\in\CC_{k}}L_{\pp,\qq}(\av^{\star},\av)=\int_{\Omega_n} d\av L_{\pp,\qq}(\av^{\star},\av)~.
\end{equation}
Design strength is preserved under unitary rotations $\R$, so logical constellations $\CC_k=\R_k\CC_0$ consisting of rotated versions of a complex spherical $\tau$-design $\CC_0$ yield a QSC whose degree distance is at least $\tau+1$.
In this way, construction of good QSCs can be accomplished by 
%analyzing the problem of 
finding well-separated spherical designs $\CC_0$ of high strength coupled with a choice of rotations $\{\R_k\}_{k=0}^{K-1}$ \spj*{(with \(\R_0\) the identity)} that permits to control the resolution $\de$ of the code constellation $\bigcup_{k}\R_{k}\CC_{0}$ while achieving high logical dimension $K$.

\vva*{
Corroborating our parameter-based analysis, we numerically compare the performance of multi- and single-mode codes using the channel fidelity \cite{schumacher_sending_1996,reimpell_iterative_2005,fletcher_optimum_2007,fletcher_channel-adapted_2008,albert_performance_2018}. 
% As expected, we observe that the additional dimensions available for multi-mode QSCs allow one to pack more quantum information into a code without sacrificing in performance.
% We observe that multi-mode QSC logical constellations are suitable for storing a large amount of quantum information (i.e., a logical space of large dimension) than single-mode cat-code constellations (see Appx.~\ref{appx:performance}).
We observe that, for quKit encodings (for \(K>2\)), even simple multi-mode constellations, such as the simplex \eqref{eq:simplex}, are able to utilize the extra dimensions efficiently and outperform single-mode constellations over a range of loss rates (see Appx.~\ref{appx:performance}).
The more non-trivial \(K=6\) Möbius-Kantor $\subset 2\{8\}3$ encoding (see Table~\ref{tab:polytope_qsc_complex}) consistently outperforms various combinations of cat codes for a wide range of energies and noise parameters.
}

\parg{CSS-based QSCs}
Concatenations of \eczoo[CSS codes]{css} \cite{calderbank_good_1995,steane_error_1996,steane_multiple-particle_1996} with the \eczoo[two-component cat code]{two-legged-cat} \cite{cochrane_macroscopically_1999}, $\CC_0=\{(+1)\}=-\CC_1$, can also be interpreted as QSCs, albeit with a weight-based notion of ladder-error protection. Such codes are actively studied \cite{cohen_dissipation-induced_2014,fukui_analog_2017,fukui_high-threshold_2018,guillaud_repetition_2019,vuillot_quantum_2019,noh_fault-tolerant_2020,chamberland_building_2022,larsen_fault-tolerant_2021,guillaud_error_2021,noh_low-overhead_2022,zhang_concatenation_2022,regent_high-performance_2022,stafford_biased_2022,lieu_candidate_2022,Gouzien_Computing_2023}, but have so far been interpreted in the framework of the outer qubit code and not in terms of underlying modal degrees of freedom.
Our interpretation parallels a standard way to construct (classical) spherical codes
by mapping binary codes to the (real) sphere 
[\citealp{conway_sphere_1999},~Sec.~2.5;~\citealp{ericson_codes_2001},~Sec.~1.2].

A $[[n,k,(d_X,d_Z)]]$ qubit CSS code is constructed from two binary linear codes with distances $d_X$ and $d_Z$, guaranteeing detection of Pauli $X$-type and $Z$-type errors with weights less than the distances, respectively.
Its codewords are equal superpositions of multi-qubit states labeled by binary strings.
Concatenation 
%with the two-component cat code 
is equivalent to mapping each binary string into a point on the $n$-sphere via the coordinate-wise antipodal mapping $0\to+1$ and $1\to-1$.
This yields an $\sph{n,2^k,\de=4d_X/n,w_{\updownarrow}=d_Z}$ QSC that detects all errors $L_{\pp,\qq}$ with Hamming weight $\Delta(\pp+\qq) < w_{\updownarrow}$ (see Appx. \ref{appx:css}).
Asymptotically good qubit CSS codes thus yield QSCs whose distances $\de,w_{\updownarrow}$ are both separated from 0 as $n\to\infty$.

\parg{X-type gates \& stabilizers}
Rotations on the $n$-sphere provide groups of $X$-type logical gates and stabilizers for QSCs.
Elements of a \textit{logical group} $\grp{G}$ permute logical constellations.
%$\CC_k$,
Elements of a \textit{stabilizer subgroup} $\grp{H}\subset\grp{G}$ permute points within each constellation, thereby leaving codewords invariant.
Rotations are realized by passive linear-optical transformations using \cite[Eq. (3.24)]{serafini_quantum_2017}.
%, yielding permutation-based logical gates.
Rotation-based gates are noise-bias preserving \cite{puri_bias-preserving_2020} in that they do not convert rotations into losses. %they can be done for any energy $\nb$.
%Such rotations are realized by passive Gaussian transformations.

For cat codes with $2p$ components, $\CC_0=\{(\zeta^{2j})\,|\,j\in\grp{\bbZ_p}\}=\zeta\CC_1$ with $\zeta=e^{\i\frac{\pi}{p}}$,
the 1D rotation $\zeta$ permutes the two constellations, while powers of $\zeta^2$ leave each constellation invariant.
These rotations generate $\grp{H}=\grp{\bbZ_{p}}\subset\grp{G}=\grp{\bbZ_{2p}}$ and are realized by 
%single-mode
transformations $\zeta^{a^{\dg} a}$ and $\zeta^{2a^{\dg} a}$.

Simplex constellations \eqref{eq:simplex} can be permuted with the $-\left(\begin{smallmatrix}1 & 0\\
0 & 1
\end{smallmatrix}\right)$ rotation and are invariant under powers of $\o\left(\begin{smallmatrix}1 & 0\\
0 & \o
\end{smallmatrix}\right)$, corresponding to the groups $\grp{\bbZ_5}\subset\grp{\bbZ_{5}}\times\grp{\bbZ_{2}}$, respectively.
%, $\CC_0=\{(\o^{\mu},\o^{2\mu}),\mu\in\bbZ_{5}\}=-\CC_0$ with $\omega=e^{\i\frac{2\pi}{5}}$,
The latter group is generated by the two-mode transformations $(-1)^{a_1^{\dg} a_1+a_2^{\dg} a_2}$ and $\o^{a_1^{\dg} a_1+2a_2^{\dg} a_2}$.

%Both groups are non-Abelian for the Möbius-Kantor and Hessian codes.
A stabilizer group for the Hessian code \eqref{eq:hessian} is $\grp{He_3}=\bra\eta,X,Z\ket$, the 27-element qutrit Pauli/$\grp{He}$isenberg group consisting of powers of $\eta$ and the $X,Z$ qutrit Pauli matrices.
Appending by the logical-$X$ rotation $-I$, where $I$ is the 3-by-3 identity, yields the logical group $\grp{He_3\times\bbZ_2}$.
These groups are realized by phase-shifters and SWAP gates.
Larger $\grp{H}\subset\grp{G}$ can be picked using the fact that all constellations form polytopes.
The largest 
% \st{, $\grp{3[3]3[3]3}\subset\grp{2[4]3[3]3}$, form}
\spj*{such groups are the 648-element and 1296-element symmetry groups of the corresponding Hessian and double-Hessian polytopes, respectively.}
These offer other ways to implement the logical-$X$ Pauli gate, but do not yield other gates.

%Other polytope codes admit similar gate constructions, with
Qudit QSCs offer larger logical-gate groups. 
The two groups are $\grp{\bbZ_2}\subset\grp{2I}$ for the 24-cell $\sph{2,5,0.382,\bra4,6,8\ket}$ real polytope code, with the former generated by the 5-by-5 matrix $-I$, and the latter the binary icosahedral group $\grp{2I}$.
%Applying the orbit-stabilizer theorem \cite{},
Since the stabilizer group acts trivially, the logical group acts on the 5 codewords as a 5D permutation representation of the icosahedral group $\grp{I}=\grp{2I/\bbZ_2}$.

CSS-based QSCs inherit logical-$X$ stabilizers (gates) by mapping each $X$-type stabilizer (logical Pauli) to a transversal linear-optical transformation via the component-wise mapping $\sigma_x\to(-1)^{a^{\dg} a}$. For example, the $\sigma_x^{\otimes 4}$ stabilizer of the $[[4,2,2]]$ \eczoo[code]{stab_4_2_2} is mapped to the joint parity $\bigotimes_{j=1}^4 (-1)^{a_j^{\dg} a_j}$.

\parg{Z-type gates \& stabilizers}
The $Z$-type ``stabilizer'' for $2p$-component cat codes is $F(a)=a^{2p}-\nb^p$, which annihilates each point in the dilated code constellation $\sqrt{\nb}\CC$.
The corresponding polynomial $F(\a)$ can be thought of as a potential on the sphere that is minimized only at the code-constellation points \cite{mirrahimi_dynamically_2014}.

Polytope QSCs can require multiple polynomials to be stabilized.
Simplex codes \eqref{eq:simplex} are stabilized by $F_1=a_1^{2}a_2^{4}-\nb^{3}$ and $F_2=a_1^{3}a_2-\nb^{2}$.
Hessian codewords \eqref{eq:hessian} are stabilized by the $F_1=a_1 a_2 a_3$, $F_2=a_1^{3}+a_2^{3}+a_3^{3}$, and $F_3=a_1^{6}+a_2^{6}+a_3^{6}-\nb^{3}/4$.
The degree of $F_{1,2}$ is lower than the code's degree distance ($d_{\updownarrow}=5$) and detectable-loss distance ($d_{\downarrow}=9$), unlike for the cat codes.
\vva{
This property makes this code similar to  degenerate stabilizer codes \eczoo{qubit_stabilizer}, i.e., codes whose check-operator weight is smaller than their distance.
}
% \spj{Please check if this is correct, I just put this here reading the reviewer comment}This \spj*{degeneracy effect is also manifest in \eczoo[QLDPC codes]{qldpc}(and more generally, in \eczoo[stabilizer]{qubit_stabilizer} and \eczoo[subsystem codes]{subsystem_qubits_into_qubits}),} which 
% admit low-weight check operators but can have larger distances.

Stabilizer polynomials commute with logical transformations $U_{\R}$ for any $\R$ in the logical group and
can be obtained by averaging ladder operators \eqref{eq:ladder} over the symmetry group of the code constellation's polytope.

Other polynomials act as logical gates on QSCs, evaluating to the same value for all points in $\CC_k$ in a way that depends on $k$.
For the cat codes, $G=a^p$ evaluates to $\pm \nb^{p/2}$ on the two codewords, respectively, yielding a logical-$Z$ gate.
The monomial $G=a_1a_2^{2}$ projects to a logical-$Z$ gate within the simplex codespace.
The smallest loss-only $Z$-gate of the Hessian code is $G_1 = a_1^3a_2^6$ or its two cyclic permutations, and
only a permutation-symmetric combination of all three operators commutes with the stabilizer group.
%, where ``perms.'' includes the two terms where $a_j$ are cyclically permuted.
%Such permutations are not necessary in the sense that
%$ a_2^6 + a_2^3 a_3^6 + a_1^6 a_3^3$.
%Each of the three terms projects to the same logical gate, but only a permutation-symmetric combination commutes with the stabilizer group $\grp{He_3}$.
%a_2^3 a_3^6 + a_1^6 a_3^3$.
A lower-degree monomial $G_2=a^{\dg}_1 a_1 a_2^3$ realizes another $Z$-gate with the help of gain operators.
Combinations $G_j + G_j^\dg$ generate logical $Z$-rotations within the $F$-annihilated subspace \cite{mirrahimi_dynamically_2014}, and have been realized for $p=2$ cat codes \cite{touzard_coherent_2018}.
%Due to detecting a large number of detectable losses,

CSS-based QSCs inherit gates/stabilizers by mapping each $Z$-type gate/stabilizer to a monomial via the component-wise mapping $\sigma_z \to a$. For example, the $[[4,2,2]]$ code's $\sigma_z\otimes\sigma_z\otimes I \otimes I$ gate is mapped to $a_1a_2$.
These codes also require stabilizers $a_j^2-\nb/n$ on each mode $j$ in order to stabilize the inner cat-code constellation.

\parg{Correcting errors}
Protection against rotation-based noise for $2p$-component cat codes is done passively using a Lindbladian whose jump operator is the $Z$-type stabilizer $F$ \cite{mirrahimi_dynamically_2014} and/or a Hamiltonian $F^{\dg} F$ \cite{goto_bifurcation-based_2016,puri_engineering_2017}.
Both techniques have been realized for $p=2$ \cite{leghtas_confining_2015,grimm_stabilization_2020}.  
General QSCs admit the same type of passive protection but require several $F_j$'s.

\vva*{
Microwave cavities coupled to superconducting circuits \cite{girvin_introduction_2023} provide a fertile ground for realizing such passive protection, and we outline how an existing superconducting circuit element called an ``ATS'' \cite{lescanne_exponential_2020} can be tuned to realize the more complicated jump operators of several QSCs (see Appx.~\ref{appx:lindblad}).
In particular, we show that a recent surface-cat concatenated-code proposal \cite{chamberland_building_2022} can be readily modified with a \(Z\)-type surface-code stabilization scheme, thereby utilizing the full power of the code against \(Z\)-type noise in exclusively passive fashion.
}

Ladder errors \eqref{eq:ladder} map the $k$th codeword \eqref{eq:coherent_states} into error states in $\text{span}\{|\av\ket,\av\in\CC_k\}$.
The stabilizer group $\grp{H}$ splits up into several irreducible representations (irreps) acting on this span. Ladder-error protection is done by measuring syndromes associated with irreps and mapping back into the codespace.
In order for correction to be possible, the stabilizer group has to be able to resolve all error spaces associated with a given error set.

The $4$-component cat-code stabilizer is the parity $(-1)^{a^{\dg} a}$. Its eigenvalues correspond to the two irreps of $\grp{H}=\grp{Z_2}$, distinguishing between no error and a single loss $a$.
This technique \cite{sun_tracking_2014} led to the first demonstration of break-even QEC using $p=2$ cat codes \cite{ofek_extending_2016}.
Similar multimode parities detect $X$-errors for CSS-based QSCs.

For the simplex code \eqref{eq:simplex}, eigenvalues of the two-mode stabilizer $\o^{a_1^{\dg} a_1 + 2a_2^{\dg} a_2}$ label the five irreps of $\grp{Z_5}$.
They allow for correction of $\{a_1,a_2,a_1a_2,a_2^2\}$, falling short of correcting all two-mode losses due to $a_1^2$ not being simultaneously correctable with $a_2$.

For the Hessian code \eqref{eq:hessian}, the transformations realizing $\grp{He_3}$ can be measured to resolve the group's 11 irreps.
The general procedure for this and other non-Abelian codes resembles that of molecular codes \cite[Sec. V.D]{albert_robust_2020}.

% Correction for cat codes is not necessary if there are errors that can bring error states back into the codespace, e.g., $a(a|\CC_0\ket)\propto|\CC_1\ket$ for $p=2$.
% As long as such errors are detected, they can be tracked without correction \cite{mirrahimi_dynamically_2014}.
% Tracking is not possible for QSCs
% %.
% %Ladder errors have to be corrected whenever a
% that contain a constellation point with a zero coordinate ($\a_j=0$ for some $j$) because the coherent state corresponding to such a zero would be %irreversibly
% annihilated by $a_j$.
% However, since a globally rotated QSC $\R\CC$ performs as well as $\CC$, constellations can be moved away from such zero coordinates.

\parg{Conclusion}
We introduce a framework for constructing quantum analogs of the classical spherical codes,
encapsulating several physically relevant quantum coding schemes for bosonic, spin, and molecular systems.
\vva*{We apply our framework to obtain multi-mode coherent-state codes based on polytopes, CSS codes, and classical codes.
These QSCs outperform previous cat-code constructions \cite{cochrane_macroscopically_1999,leghtas_hardware-efficient_2013,denys_2t-qutrit_2022} both in terms of code parameters and a numerical performance comparison.
We show how passive protection of several instances of these QSCs can be realized in microwave cavities.
}

There are many other ways of constructing spherical codes, e.g., 
as \eczoo[group-orbit codes]{slepian_group}
\cite{slepian_group_1968,loeliger_signal_1991,mittelholzer_group_1996}, as spherical embeddings of association schemes \cite{ericson_codes_2001},
through computer searches \cite{sloane_spherical_2000,ballinger_experimental_2009}, and many others \cite{cameron_strongly_1978,conway_sphere_1999,ericson_codes_2001,waldron_introduction_2018}, as well as ways of constructing spherical designs
\cite{bajnok_construction_1991,reznick_constructions_1995,xiang_explicit_2022}.
%and that our framework should be extendable to other continuous configuration spaces, 
As such, we anticipate that this work will pave the way for many novel, well-protected, and experimentally feasible logical qubits.

\section*{Data availability statement}

\textsc{Mathematica} notebooks generated during the current study are available from the corresponding author on reasonable request.
Further details and references about spherical codes described in this manuscript are available at the Error-correction Zoo website at \href{http://errorcorrectionzoo.org/}{http://errorcorrectionzoo.org} and the corresponding repositories at \href{http://github.com/errorcorrectionzoo/eczoo_data/}{http://github.com/errorcorrectionzoo}.

\begin{acknowledgments}
\vva*{We thank
Francesco Arzani, 
Ansgar Burchards,
Jonathan Conrad,
Aurélie Denys,
Philippe Faist,
Michael Gullans,
Wenhao He,
Liang Jiang,
Greg Kuperberg,
Gideon Lee,
Anthony Leverrier,
Pavel Panteleev,
Shraddha Singh,
and Guo (Jerry) Zheng 
for helpful discussions.
VVA is especially grateful to Kyungjoo Noh for discussion about realizations of these codes.
}

This work is supported in part by NSF QLCI grant OMA-2120757 and NSF grants CCF-2110113 (NSF-BSF) and CCF-2104489.
JTI thanks the Joint Quantum Institute at the University of Maryland for support through a JQI fellowship.
Our figures were drawn using \textsc{Mathematica 13} following the prescription of Ref. \cite{coxeter_portraits_1992}.
Contributions to this work by NIST, an agency of the US government, are not subject to US copyright.
Any mention of commercial products does not indicate endorsement by NIST.
VVA thanks Ryhor Kandratsenia and Olga Albert for providing daycare support throughout this work.

\end{acknowledgments}

%\toc

\newpage
~
\newpage
\onecolumngrid
\appendix
\renewcommand{\tocname}{Appendices}

%\pagebreak
\section{Real-polytope QSCs}
\label{appx:poly_real}

%%%%%%%%%%%%%%%%%%%%%%%%%%%%%%%%%%%%%%%%%%%%%%%%%%%%%%%%
\begin{table}
\begin{tabular}{ccccccc}
\toprule
logical constellation & code constellation & $n$ & $K$ & $\langle t_\downarrow, d_{\updownarrow}, d_\downarrow\rangle$ & $d_{E}$ & related code \tabularnewline

\midrule
line segment & $2K$-gon & $1$ & $K$ & $\langle 2,2,2
\rangle$ & $4\sin^{2}\frac{\pi}{2K}$ & two-component cat qu$K$it\tabularnewline
 & icosahedron & $2$ & $6$ & $2$ & $1.106$ & \tabularnewline
 & dodecahedron & $2$ & $10$ & $2$ & $0.509$ & \tabularnewline
 & 24-cell & $2$ & $12$ & $2$ & $1.000$ & $\grp{\bbZ_2}\subset\grp{2T}$ group-GKP \tabularnewline
 & 288-cell & $2$ & $24$ & $2$ & $0.586$ & $\grp{\bbZ_2}\subset\grp{2O}$ group-GKP \tabularnewline
 & hyper-icosahedron & $2$ & $60$ & $2$ & $0.382$ & $\grp{\bbZ_2}\subset\grp{2I}$ group-GKP \tabularnewline
 & hyper-dodecahedron & $2$ & $300$ & $2$ & $0.073$ & \tabularnewline
 % & $D$-bisimplex & $\left\lceil D/2\right\rceil $ & $D+1$ & $2$ & $2-2/D$ & \tabularnewline
 & $D$-orthoplex & $\left\lceil D/2\right\rceil $ & $D$ & $\bra1,2,2\ket$ & $2.000$ & $D=4$: $\grp{\bbZ_2}\subset \grp{Q}$ group-GKP  \tabularnewline
 & $D$-cube & $\left\lceil D/2\right\rceil $ & $2^{D-1}$ & $2$ & $4/D$ & \tabularnewline
\midrule
$p$-gon & $Kp$-gon & $1$ & $K$ & $\langle p,p,p
\rangle$ & $4\sin^{2}\frac{\pi}{Kp}$ & $p$-component cat qu$K$it\tabularnewline
\midrule
%tetrahedron & cube & $2$ & $2$ & $3$ & $1.333$ & \tabularnewline

% \shubham{tetrahedron} & cube & $2$ & $2$ & $3$ & $4/3$ & \tabularnewline

tetrahedron & dodecahedron & $2$ & $5$ & $3$ & $0.509$ & \tabularnewline
\midrule
octahedron & 5-octahedron & $2$ & $5$ & $4$ & $0.382$ & \tabularnewline
\midrule
\vva*{icosahedron} & \vva*{2-icosahedron} & $2$ & $2$ & $\langle 4,6,6 \rangle$ & $0.211$ & \tabularnewline
\midrule
hyper-tetrahedron & hyper-dodecahedron & $2$ & $120$ & $3$ & $0.073$ & \tabularnewline
\midrule
%& hyper-cube & $2$ & $2$ & $4$ & $1.000$ & \tabularnewline
hyper-octahedron & 24-cell & $2$ & $3$ & $\langle 2,4,4 \rangle$ & $1.000$ & $\grp{Q}\subset\grp{2T}$ group-GKP, $\grp{2T}$-qutrit\tabularnewline
& 288-cell & $2$ & $6$ & $\langle 2,4,4 \rangle$ & $0.586$ & $\grp{Q}\subset\grp{2O}$ group-GKP \tabularnewline
& hyper-icosahedron & $2$ & $15$ & $4$ & $0.382$ & $\grp{Q}\subset\grp{2I}$ group-GKP \tabularnewline
& hyper-dodecahedron & $2$ & $75$ & $4$ & $0.073$ & \tabularnewline
\midrule
hyper-cube, -octahedron & 24-cell & $2$ & $2$ & $\langle 2,4,4 \rangle$ & $1.000$ & \tabularnewline
\midrule
24-cell & 288-cell & $2$ & $2$ & $\langle 5,6,12 \rangle$ & $0.586$ & $\grp{2T}\subset\grp{2O}$ group-GKP \tabularnewline
& hyper-icosahedron & $2$ & $5$ & $\langle 4,6,8 \rangle$ & $0.382$ & $\grp{2T}\subset\grp{2I}$ group-GKP \tabularnewline
& hyper-dodecahedron & $2$ & $25$ & $6$ & $0.073$ & \tabularnewline
\midrule
hyper-icosahedron & hyper-dodecahedron & $2$ & $5$ & $12$ & $0.073$ & \tabularnewline
\midrule
$D$-simplex & $D$-bisimplex & $\left\lceil D/2\right\rceil $ & $2$ & $\langle 2,3,3 \rangle$ & $2-2/D$ & \tabularnewline
$(2^r-1)$-simplex & $(2^r-1)$-cube & $2^{r-1}$ & $2^{2^r-r-1}$ & $3$ & $4/(2^r-1)$ & shortened Hadamard \tabularnewline
\midrule
$D$-demicube & $D$-cube & $\left\lceil D/2\right\rceil $ & $2$ & $\min(4,D)$ & $4/D$ & single parity-check\tabularnewline
\midrule
$2^r$-orthoplex & $2^r$-cube & $2^{r-1}$ & $2^{2^r-r-1}$ & $4$ & $2^{2-r}$ & augmented Hadamard \tabularnewline
\bottomrule
\end{tabular}
\caption{\label{tab:polytope_qsc} QSCs whose logical and code constellations both make up the vertices
of a real polytope; $D\protect\geq2$ corresponds to spatial dimension, and the parameter $r \geq 2$.
%\SJComment{Did we ever check the $d_\downarrow$, $t_\downarrow$ for hyper-icosahedron code? Would be interesting to compare that with the giant complex polytopes with same $d_\updownarrow$. I can run it if you have the code logicals coordinates}.
%The first (second) column lists the polytope whose vertices make up the logical (code) constellation $\CC_k$ ($\CC$).
}
\end{table}
%%%%%%%%%%%%%%%%%%%%%%%%%%%%%%%%%%%%%%%%%%%%%%%%%%%%%%%%

Logical constellations $\CC_k$ of a real polytope QSC form the vertices of a real polytope. The figure that results from the union of all logical polytopes is called a \textit{polytope compound}, and its vertices form the code constellation $\CC$.
Polytope QSCs can thus be constructed from established polytope compounds.
%, and their code properties inferred from those of the corresponding polytopes.

% \spj{Are you sure all of these compound structures are fully classified in 4D? Mccullen paper abstract seems to suggest that Coxeter classification was not complete and it still might not be after his additional compounds. Moreover, does the table really have all of those classified compounds? Did you say 'believe' because we can't tell if those classified regular compounds are convex?} 
Regular real polytope compounds have been classified in three \cite{skilling_uniform_1976} and four \cite{coxeter_regular_1973,mcmullen_new_2018} dimensions.
%, and the non-regular case remains to be completed.
\vva{
We collect all QSCs whose logical and code constellations each form a \textit{single} regular real polytope in Table \ref{tab:polytope_qsc}.
We leave to future work QSCs made up of polytope compounds whose code constellation forms vertices of \textit{multiple} regular polytopes \cite{skilling_uniform_1976}\cite[Table VII]{coxeter_regular_1973}\cite[Sec.~10]{mcmullen_new_2018} as well as recently discovered variations of compounds with the same parameters \cite[Sec.~10]{mcmullen_new_2018}.
We include a few QSCs constructed from notable non-regular polytopes.
}
All polytopes used in our constructions are listed in Table \ref{tab:real_polytopes}.

The first column of the table lists the polytope whose vertices make up the logical constellations $\CC_k$. All $\CC_k$ make up the same polytope for every code, with the exception being the ``hyper-cube, -octahedron'' code, in which $\CC_0$ ($\CC_1$) makes up the vertices of a hyper-cube (hyper-octahedron).

Since the $n$-sphere is complex while the polytopes are real, we have to embed the polytopes into the sphere.
For even dimension $D$, the standard method of doing this is via the mapping
\begin{equation}
    \bbR^{D}\ni (x_1,x_2,\cdots,x_{D}) \to (x_1 + \i x_2,x_3 + \i x_4, \cdots, x_{D-1}+\i x_{D}) \in \bbC^{D/2}~.
\end{equation}
Other mappings can be obtained by permuting the real coordinates.
For odd $D$, one has to embed the polytope into $D+1$ dimensions and then apply a mapping like the one above.
Convenient coordinates exist for polytopes embedded in higher dimensions, e.g., vertices of a $D$-simplex have coordinates $(1,1,\cdots 1,-D)\in\bbR^{D+1}$ and permutations thereof \cite[Sec. 1.5]{ericson_codes_2001}.
Mappings into higher-dimensional spaces can also be used, e.g., the $2p$-component cat-code constellation can be mapped into 
%\SJComment{1) Have you defined the $i$ in $i^\ell$? Its not iota anymore but the 2p-th root of unity. 2) Why not generalise it to Kp-components? And then it'll be the same as the last row of Table \ref{tab:polytope_qsc_complex} which we can then remove from the complex table.} 
$\CC=\{\zeta^{j}\av,j\in\grp{\bbZ_{2p}}\}$ for any $n$-dimensional unit vector $\av$.
If one prefers to use real-valued vertices, then $\bbR^D$ can be directly embedded into $\bbC^{D}$.

The parameters $t_{\downarrow},d_{\downarrow}$ can depend on which of the above mappings one uses; we calculate them numerically by evaluating Eq. \eqref{eq:ladder_kl} for a given $\CC$.
A mapping-independent lower bound on the degree distance $d_{\updownarrow}$ can be obtained from the strength of the design formed by the logical polytopes.
Real polytope vertices can form (real) spherical designs \cite{delsarte_spherical_1977}, which are convertible into complex spherical designs via \cite[Lemma 3.3]{roy_complex_2014}.
The design strengths $\tau$ of $D$-dimensional polytope vertices are listed in Table \ref{tab:real_polytopes}, column 5, yielding $d_{\updownarrow}\geq\tau+1$ for a code consisting of such polytopes.
This bound appears to be tight for real polytopes and holds as long as the polytope formed by $\CC$ is the same dimension as those formed by each $\CC_k$.
Otherwise, the logical polytopes will not share a common sphere on which their vertices form designs.
%such polytopes lie on the same $D$-dimensional real sphere embedded inside the complex $n$-sphere.
An exception to this restriction is for $\CC_k$ that are 1D line segments and is due to the fact that any pair of segments shares a common circle. 
The degree distance of a QSC consisting of segments is thus at least two.
%because any two logical constellations form a one-design on the circle defined by their four vertices.

%%%%%%%%%%%%%%%%%%%%%%%%%%%%%%%%%%%%%%%%%%%%%%%%%%%%%%%%
\begin{table}

\begin{tabular}{cccccccc}
\toprule
polytope & dim & Schlafli/Coxeter & vertices & design & $d_{E}$& $d_{E}$ (numerical) & reference\tabularnewline
\midrule
line segment & $1$ & $\{\phantom{2}\}$ & $2$ & $1$ & $4$ & $4.000$ &\tabularnewline
\midrule
triangle & $2$ & $\{3\}$ & $3$ & $2$ & $3$ & $3.000$ &\tabularnewline
square & $2$  & $\{4\}$ & $4$ & $3$ & $2$ & $2.000$ &\tabularnewline
pentagon & $2$ & $\{5\}$ & $5$& $4$ & $\sqrt{5}(1-\varphi)$ & $1.382$ &\tabularnewline
$\vdots$ & $\vdots$ & $\vdots$ & $\vdots$ & $\vdots$& $\vdots$& $\vdots$ &\tabularnewline
$p$-gon  & $2$ & $\{p\}$ & $p$& $p-1$  & $4\sin^{2}\frac{\pi}{p}$ &  &\tabularnewline
\midrule
tetrahedron & $3$ & $\{3,3\}$ & $4$ & $2$ & $8/3$ & $2.667$ &\tabularnewline
octahedron & $3$ & $\{3,4\}$ & $6$  & $3$ & $2$ & $2.000$ &\tabularnewline
cube & $3$ & $\{4,3\}$ & $8$  & $3$ & $4/3$ & $1.333$ &\tabularnewline
%stella octangula
icosahedron & $3$ & $\{3,5\}$ & $12$  & $5$ & $4/(1+\varphi^2)$ & $1.106$ &\tabularnewline
dodecahedron & $3$ & $\{5,3\}$ & $20$ & $5$ & $2-2\sqrt{5}/3$ & $0.509$ &\tabularnewline
\vva*{\textit{2-icosahedron}} & $3$ & $\mathit{\beta\{3,4\}}$ & $24$ & $5$ & $2(1-\varphi)^2/(1+\varphi^2)$ & $0.2111$ & \cite{skilling_uniform_1976}\tabularnewline
\textit{5-octahedron} & $3$ & $\mathit{[5\{3,4\}]2\{3,5\}}$ & $30$ & $5$ & $(1-\varphi)^2$ & $0.382$ & \cite{skilling_uniform_1976}\tabularnewline
% Orbit of I group, see 10.2969/aspm/02410309
\midrule
hyper-tetrahedron & $4$ & $\{3,3,3\}$ & $5$ & $2$ & $5/2$ & $2.500$ & \cite{mamone_orientational_2010} \tabularnewline
hyper-octahedron & $4$ & $\{3,3,4\}$ & $8$ & $3$ & $2$ & $2.000$ & \cite{mamone_orientational_2010} \tabularnewline
hyper-cube & $4$ & $\{4,3,3\}$ & $16$ & $3$ & $1$ & $1.000$ & \cite{mamone_orientational_2010} \tabularnewline
24-cell & $4$ & $\{3,4,3\}$ & $24$ & $5$ & $1$ & $1.000$ & \eczoo{24cell} \cite{mamone_orientational_2010} \tabularnewline
\textit{288-cell} & $4$ & \textit{o3m4m3o} & $48$ & $7$ & $2-\sqrt{2}$ & $0.586$ & \eczoo{disphenoidal288cell} \cite{klitzing_polytopes_2023} \tabularnewline
hyper-icosahedron & $4$ & $\{3,3,5\}$ & $120$ & $11$ & $(1-\varphi)^2$ & $0.382$ & \eczoo{600cell} \cite{mamone_orientational_2010}\tabularnewline
hyper-dodecahedron & $4$ & $\{5,3,3\}$ & $600$ & $11$ & $(7-3 \sqrt{5})/4$ & $0.073$ & \eczoo{120cell} \cite{mamone_orientational_2010} \tabularnewline
\midrule
$D$-simplex & $D$ & $\{3^{D-1}\}$ & $D+1$  & $2$ & $2+2/D$ &  & \eczoo{simplex_spherical}\tabularnewline
$D$\textit{-bisimplex} & $D$ & $\mathit{[2\{3^{D-1}\}]}$ & $2(D+1)$ & $2$ & $2-2/D$ &  & \cite{klitzing_polytopes_2023} \tabularnewline
$D$-orthoplex & $D$ & $\{3^{D-2},4\}$ & $2D$  & $3$ & $2$ & $2.000$ & \eczoo{biorthogonal}\tabularnewline
$D$\textit{-demicube} & $D$ & $\mathit{\{3^{1,D-3,1}\}}$ & $2^{D-1}$  & $\min(3,D-1)$ & $8/D$ &  & \cite{klitzing_polytopes_2023} \tabularnewline
$D$-cube & $D$ & $\{4,3^{D-2}\}$ & $2^{D}$ & $3$ & $4/D$ &  &\tabularnewline
\bottomrule
\end{tabular}\caption{\label{tab:real_polytopes}Polytope data used to construct QSCs in Table \ref{tab:polytope_qsc}. Non-italicised polytopes make up the convex regular polytopes in real dimension $D$.  \vva*{$\varphi=\frac{1+\sqrt{5}}{2}$ is the golden ratio.}}

\end{table}
%%%%%%%%%%%%%%%%%%%%%%%%%%%%%%%%%%%%%%%%%%%%%%%%%%%%%%%%

Points on the real 4D sphere are in one-to-one correspondence with quaternions, which in turn parameterize the group $\grp{SU(2)}$ \cite{hanson_visualizing_2005}.
Vertices of the hyper-octahedron, 24-cell, (disphenoidal) 288-cell, and hyper-icosahedron correspond to quaternions forming the quaternion $\grp{Q}$, binary tetrahedral $\grp{2T}$, binary octahedral $\grp{2O}$, and binary icosahedral $\grp{2I}$ subgroups, respectively.
Polytope QSCs consisting of such polytopes thus are related to $\grp{SU(2)}$ \eczoo[group-GKP codes]{group_gkp} \cite{albert_robust_2020}.
The $\grp{2T}$-qutrit code \cite{denys_2t-qutrit_2022} is similarly related to the $\grp{Q}\subset\grp{2T}\subset\grp{SU(2)}$ group-GKP code, but the idea of using groups this way is limited to two modes because spheres in higher dimensions no longer correspond to groups.

An $[n,k]$ \eczoo[binary linear]{binary_linear} code $C$ can be converted into a QSC by taking logical constellations to be cosets of $C$ in $\bbF_2^n$ under the antipodal mapping.
The table lists QSCs arising this way from the \eczoo[Hadamard]{hadamard} and \eczoo[single parity-check]{parity_check} codes.
These codes all have non-trivial $d_{\updownarrow}$ because the cosets correspond to known polytope compounds when embedded into the sphere \cite[pg. 287]{coxeter_regular_1991}.

\section{Complex-polytope QSCs}
\label{appx:poly_complex}

\begin{table}
\begin{tabular}{ccccccc}
\toprule 
logical const-n & code const-n & $n$ & $K$ & $\langle t_\downarrow, d_{\updownarrow}, d_\downarrow\rangle$ & $d_{E}$ & related code \tabularnewline

\midrule

Möbius-Kantor  & $2\{6\}3$  & $2$ & $2$ & $\langle 3,4,6 \rangle$ & $0.845$ & \tabularnewline
  & $3\{4\}3$  & $2$ & $3$ & $\langle 3,4,4 \rangle$ & $1.000$ & $\grp{Q}\subset\grp{2T}$ group-GKP  \tabularnewline
  & $2\{8\}3$  & $2$ & $6$ & $\langle 3,4,4 \rangle$ & $0.367$ & \tabularnewline
\midrule
$(2,4)$-orthoplex & $4\{3\}4$  & $2$ & $3$ & $\langle 2,4,4 \rangle$ & $1.000$ & $\grp{Q}\subset\grp{2T}$ group-GKP, $\grp{2T}$-qutrit \tabularnewline
\midrule
$3\{6\}2$ & $[2~3\{6\}2]$  & $2$ & $2$ & $\langle 4,4,4 \rangle$ & $0.211$ & \tabularnewline
\midrule
$4\{3\}4$ & $2\{6\}4$  & $2$ & $2$ & $\langle 5,6,12 \rangle$ & $0.586$ & $\grp{2T}\subset\grp{2O}$ group-GKP  \tabularnewline
\midrule
$3\{4\}3$ & $2\{8\}3$  & $2$ & $2$ & $\langle 3,6,12 \rangle$ & $0.367$ & \tabularnewline
\midrule
$2\{6\}4$ & $[2~2\{6\}4]$  & $2$ & $2$ & $\langle 4,8,8 \rangle$ & $0.268$ & \tabularnewline
\midrule
$3\{5\}3$ & $2\{10\}3$  & $2$ & $2$ & $\langle 9,12,30 \rangle$ & $0.132$ & \tabularnewline
\midrule
$5\{3\}5$ & $2\{6\}5$  & $2$ & $2$ & $\langle 11,12,30 \rangle$ & $0.098$ & \tabularnewline
 & $3\{4\}5$  & $2$ & $3$ & $\langle 11, 12,20 \rangle$ & $0.044$ & \tabularnewline
\midrule
$(3,3)$-orthoplex & rectified Hessian & $3$ & $8$ & $\langle 2,3,3 \rangle$ & $1.000$ & \tabularnewline
\midrule
$(3,6)$-orthoplex & rectified Hessian & $3$ & $4$ & $\langle 2,4,6 \rangle$ & $1.000$ & \tabularnewline
\midrule
Hessian & double Hessian & $3$ & $2$ & $\langle 4,5,9 \rangle$ & $1.000$ & \tabularnewline
\midrule
Witting & double Witting & $4$ & $2$ & $\langle 6,8,12 \rangle$ & $0.586$ & Clifford group-orbit \tabularnewline
\midrule
$(1,m)$-cube & $(n,m)$-cube & $n$ & $m^{n-1}$ & $\bra1,2,m\ket$ & $\frac{4}{n}\sin^2\frac{\pi}{m}$ & \tabularnewline 
\midrule
 %$(1,4)$-orthoplex  & $(n,4)$-orthoplex  & $n$ & $2n$ & $\bra1,2,2\ket$ & $2.000$ & \tabularnewline 
  $(1,m)$-orthoplex  & $(n,m)$-orthoplex  & $n$ & $n$ & $\bra1,2,m\ket$ & $\min(2,4\sin^2\frac{\pi}{m})$ & \tabularnewline 
% \midrule
%$(n,m)$-orthoplex & $(n,Km)$-orthoplex   & $n$ & $K$ & $min(4,m)^\star$ & $\min(2,4\sin^2(\frac{\pi}{Km}))$ & \tabularnewline
\bottomrule
\end{tabular}
\caption{\label{tab:polytope_qsc_complex} QSCs whose logical and code constellations both make up the vertices
of a non-real complex polytope; $n\protect\geq1$ corresponds to complex dimension. $d_\downarrow=m$ for the $(n, m)$-cube/orthoplex codes are conjectured based on numerical results.}
\end{table}

Complex polytopes are polytopes whose vertices are complex.
As with real polytopes, there are a myriad polygons in the two complex dimensions, a handful of special polytopes in a few of the higher dimensions, and only two infinite families of non-real complex polytopes present in any dimension.

The two families are straightforward complex generalizations of the cube and orthoplex, respectively.
A simple set of vertices of a real $D$-dimensional cube consists of $2^D$ vectors with coordinates $\pm1$. 
The vertices of the \textit{complex} $(n,m)$\textit{-cube} (a.k.a.~$\gamma_n^m$) consist of $m^n$ complex vectors of dimension $n$ with $m$th roots of unity at each coordinate.
%The ordinary hyper cube ($\gamma_n$) and hyper orthoplex ($\beta_n$) can be generalised to the complex cube ($\gamma_n^m$) and the complex orthoplex ($\beta_n^m$) where the new parameter $m$ signifies the use of $m-$edges comprised of $m$ roots of unity instead of the $2-$edges comprised of $\pm 1$ in the real case. 
A similar generalization holds for the $(n,m)$\textit{-orthoplex} (a.k.a.~$\beta_n^m$), whose $mn$ coordinates are $n$-dimensional vectors whose single nonzero entry is an $m$th root of unity.
%We deviate from the traditional notation for hyper-cubes ($\gamma_n^m$) and hyper-orthoplexes ($\beta_n^m$) and use $(n,m)$-cube/orthoplex to denote the corresponding polytopes. There are a few special polytopes in $3$ and $4$ complex dimensions and a large amount of complex polygons in $2$ dimensions which do not fall into the cube or orthoplex families.

A union of complex polytopes sharing a common center forms a \textit{complex polytope compound}.
%Analogous to real compounds, 
Complex compounds yield complex QSCs whose code constellations are formed by the vertices of the compound and whose logical constellations are formed by the vertices of the participating polytopes.
%Similar to real polytope codes, logical constellations $\CC_k$ that form the vertices of a complex polytope can unite to form code \textit{complex polytope compounds} with vertices $\CC$. 
%All regular complex polytopes have been classified in all dimensions \cite{shephard_regular_1952}, but 
Complex compounds have not been as thoroughly studied as their real counterparts, and most of our codes come from the handful of constructions from Refs. \cite{coxeter_coordinates_1997, coxeter_reciprocating_1997, coxeter_regular_1991}.
In Table \ref{tab:polytope_qsc_complex}, we collect the complex polytope QSCs that are the most interesting for a comparative study with the real polytope codes.
All the polytopes used in our constructions are listed in Table \ref{tab:complex_polytopes}.
%A more extensive list of complex polytope QSCs is left to a follow-up work. 
%All of these compound polytopes, with the exceptions of ``double Witting'' and ``$[2~3\{6\}2]$'' code constellations, are extracted from existing polytope compound constructions \cite{} and adapted to QSCs. 

Complex polygons yield several interesting QSCs not available in the real case.
We mentioned already in the main text that multiple complex polytopes can reduce to the same real polytope when mapped into the reals. 
As another example, compounds consisting of $5\{3\}5$ polygonal code constellations have exceptional loss detection capabilities, with $d_\downarrow$ as high as 30, but suffer from low resolution.
There are many more polygons, and we leave a more extensive list of complex polytope QSCs to a follow-up work. 
%Other complex polygons also demonstrate a similar trade-off in the resolution and ladder distance. 
%Some of these polytopes demonstrate exceptional properties when adapted to QSCs and are listed in Table \ref{tab:complex_polytopes}.

Complex polytopes also offer interesting many-mode alternatives to cat codes.
The tensor product of $n$ single-mode $4$-component cat codes
%, each housing $2$ logical states with code constellations of size $2$, 
is an $\sph{n,2^n,2/n,\bra 2, 2, 2\ket}$ QSC whose code constellation can be thought of as an $(n, 4)$-cube, constructed as a Kronecker product of $n$ $(1, 4)$-cubes.
The resolution of this code decreases as order $O(1/n)$, meaning that a constant energy per mode (usually picked to be $\nb/n\approx2$ \cite{chamberland_building_2022,ofek_extending_2016}) is required in order to be able to resolve codewords without substantial intrinsic memory error.
On the other hand, the $(n, 4)$-orthoplex $\sph{n,n,2.0,\bra1, 2, 4\ket}$ QSC, whose logical constellations are $(1, 4)$-orthoplexes, maintains \textit{constant} resolution and has extra loss detection at the expense of a linear increase in the codespace dimension and no loss correction. 
%We did not find analogous codes using real polytopes, and 
It is an interesting open problem to find a QSC with $K=O(n)$ that can correct one or more losses.
%{\color{red} This feature is not found in any real polytope code. Are we sure? $D$-orthoplex code does this too, no?}
% This structure yields a non-polytope QSC with parameters $\sph{n,K^n,\frac{4}{n}\sin^2(\frac{\pi}{Km}),\bra m, m, m\ket}$. This ladder distance is significantly better than the $(n, Km)-$cube code in Table \ref{tab:polytope_qsc_complex}, but comes at the cost of a worse resolution $d_E$. 

\begin{table}

\begin{tabular}{cccccccc}
\toprule 
polytope & dim & Schlafli/Coxeter & vertices & design & $d_{E}$& $d_{E}$ (numerical) & reference\tabularnewline\midrule
% $(1,m)$-cube & $1$  & $\{m\}$
% & $m$  & $m-1^\star$ & $4\sin^2\frac{\pi}{m}$ & 
% \tabularnewline\midrule

Möbius-Kantor & $2$  & $3\{3\}3$    & $8$   & $3$  & $2$ & $2.000$ & \cite{shephard_regular_1952, coxeter_coordinates_1997} \tabularnewline
& $2$  & $2\{6\}3$    & $16$  & $3$  & $2-2/\sqrt{3}$ & $0.845$ & \cite{shephard_regular_1952, coxeter_coordinates_1997} \tabularnewline
& $2$  & $3\{4\}3$    & $24$  & $5$  & $1$ & $1.000$ & \cite{shephard_regular_1952, coxeter_coordinates_1997} \tabularnewline
& $2$  & $4\{3\}4$    & $24$  & $5$  & $1$ & $1.000$ & \cite{shephard_regular_1952, coxeter_coordinates_1997} \tabularnewline
& $2$  & $3\{6\}2$    & $24$  & $3$  & $(3-\sqrt{3})/2$    & $0.634$ & \cite{shephard_regular_1952, coxeter_coordinates_1997} \tabularnewline
& $2$  & $2\{6\}4$    & $48$  & $7$  & $2-\sqrt{2}$ & $0.586$ & \cite{shephard_regular_1952, coxeter_coordinates_1997} \tabularnewline
& $2$  & $2\{8\}3$    & $48$  & $5$  & $2-2\sqrt{2/3}$ & $0.367$ & \cite{shephard_regular_1952, coxeter_coordinates_1997} \tabularnewline
& $2$  & $\mathit{[2~3\{6\}2]}$ & $48$  & $3$  & $2(1-\varphi)^2/(1+\varphi^2)$ & $0.211$ & \tabularnewline
& $2$  & $\mathit{[2~2\{6\}4]}$ & $96$  & $7$  & $2-\sqrt{3}$ & $0.268$ & \tabularnewline
& $2$  & $3\{5\}3$    & $120$ & $11$ &$(1-\varphi)^2$& $0.382$ & \cite{shephard_regular_1952, coxeter_coordinates_1997} \tabularnewline
& $2$  & $5\{3\}5$    & $120$ & $11$ &$(1-\varphi)^2$& $0.382$ & \cite{shephard_regular_1952, coxeter_coordinates_1997} \tabularnewline
& $2$  & $2\{10\}3$   & $240$& $11$ & $2-\sqrt{2(3+\sqrt{5})/3}$ & $0.132$ & \cite{shephard_regular_1952, coxeter_coordinates_1997} \tabularnewline
& $2$  & $2\{6\}5$    & $240$ & $11$ & $2-\sqrt{\varphi\sqrt{5}}$ & $0.098$ & \cite{shephard_regular_1952, coxeter_coordinates_1997} \tabularnewline
& $2$  & $3\{4\}5$ & $360$& $11$ & $4 \sin^2(\pi/30)$ & $0.044$ & \cite{shephard_regular_1952, coxeter_coordinates_1997} \tabularnewline

\midrule

Hessian & $3$ & $3\{3\}3\{3\}3$% or $2_{21}$
& $27$ & $4$ & $3/2$ & $1.500$& \eczoo{hessian_polyhedron} \cite{shephard_regular_1952}\tabularnewline
double Hessian & $3$ & $2\{4\}3\{3\}3$ & $54$ & $4$ & $1$ & $1.000$& \cite{coxeter_reciprocating_1997}\tabularnewline
rectified Hessian & $3$ & $3\{3\}3\{4\}2$ & $72$ & $5$ & $1$ & $1.000$& \eczoo{rect_hessian_polyhedron} \cite{coxeter_regular_1991}\tabularnewline
\midrule
Witting & $4$ & $3\{3\}3\{3\}3\{3\}3$ & $240$ & $7$ & $1$ & $1.000$& \eczoo{witting_polytope} \cite{shephard_regular_1952}\tabularnewline
\textit{double Witting} & $4$ & $\textit{[2~3\{3\}3\{3\}3\{3\}3]}$ & $480$ & $7$ & $2-\sqrt{2}$ & $0.586$& \tabularnewline
\midrule
$(n,m)$-cube & $n$ & $m\{4\}2\{3\}\cdots2\{3\}2$ & $m^n$ & $\min(3,m-1)$ & $\frac{4}{n}\sin^2\frac{\pi}{m}$ & & \cite{shephard_regular_1952}\tabularnewline
$(n,m)$-orthoplex & $n$ & $2\{3\}2\{3\}\cdots2\{4\}m$ & $nm$ & $\min(3,m-1)$ & $\min(2,4\sin^2\frac{\pi}{m})$ & & \cite{shephard_regular_1952}\tabularnewline
\bottomrule
\end{tabular}\caption{\label{tab:complex_polytopes} Non-real polytope data used to construct QSCs in Table \ref{tab:polytope_qsc_complex}. Italicised polytopes are not regular. \vva*{$\varphi=\frac{1+\sqrt{5}}{2}$ is the golden ratio.}}

\end{table}
\section{CSS-based QSCs}
\label{appx:css}

The antipodal mapping converts binary strings $\bb=(b_{1},b_{2},\cdots,b_{n})$ labeling $n$-qubit states into $n$-mode coherent states normalized to an energy of unity,
\begin{equation}
    \av_{\bb}=\left((-1)^{b_{1}},(-1)^{b_{2}},\cdots,(-1)^{b_{n}}\right)/\sqrt{n}~.
\end{equation}
Using \cite[Thm. 7.3]{bruss_lectures_2007}, there exists a basis of codewords for an $[[n,k,(d_X,d_Z)]]$ CSS code that is labeled by length-$k$ binary strings $\ll$ and that is expressed in terms of $\grp{C_{Z}^\perp}$, the dual of one of the underlying binary linear codes.
Applying the antipodal mapping to the $\ll$th element of such a basis yields a codeword for the corresponding QSC,
\begin{equation}
    |\overline{\ll}\ket\sim\frac{1}{\sqrt{|\grp{C_{Z}^{\perp}}|}}\sum_{\cc\in\grp{C_{Z}^{\perp}}}|\sqrt{\nb}~\av_{\ll+\cc}\ket~.
\end{equation}

\parg{Phase-flip errors}
%Recall that lowering operators are eigenstates of coherent states, $a_j|\av\ket=\a_j|\av\ket$.
Using Eq. \eqref{eq:ladder_kl}, the projection of a general ladder error acting a subset of modes $\SS$ into the QSC codespace is equivalent to a $Z$-type error,
\begin{equation}\label{eq:projected}
    L_{\pp,\qq}^{(\SS)}=\prod_{j\in\SS}a_{j}^{\dg p_{j}}a_{j}^{q_{j}}\quad\quad\to\quad\quad\left(\frac{\nb}{n}\right)^{|\pp+\qq|/2}\prod_{j\in\SS}Z_{j}^{p_{j}+q_{j}}~,
\end{equation}
where we define $Z_j|\sqrt{\nb}\av_{\bb}\ket = (-1)^{b_j}|\sqrt{\nb}\av_{\bb}\ket$.
As long as the support size of the region $\SS$ is less than $d_Z$, the distance of $\grp{C_{Z}}$, the properties of CSS codes can be used to show that the above error is detectable.
This means that any ladder error with Hamming weight $\Delta(\pp+\qq)<d_Z$ is detectable.

%Since $Z_{j}^{2}=1$, we have that $p_{j}+q_{j}=1$ modulo $2$ yields $Z_{j}$, while $p_{j}+q_{j}=0$ modulo 2 yields one on the codespace. Therefore, the resulting projected operator \eqref{eq:projected} is supported on no more than $\Delta(\SS)$ modes.

\parg{Bit-flip errors}
The squared Euclidean distance between two code constellation elements $\av_{\bb}$ and $\av_{\cc}$ can be expressed in terms of the Hamming distance $\Delta(\bb,\cc)$ between their corresponding binary strings,
\begin{subequations}
\begin{align}
    \left\Vert \av_{\bb}-\av_{\cc}\right\Vert ^{2}
    &=2-2\av_{\bb}\cdot\av_{\cc}\\
    &=2-\frac{2}{n}\sum_{j=1}^{n}(-1)^{b_{j}+c_{j}}\\
    &=2-\frac{2}{n}\sum_{j=1}^{n}+[n-\Delta(\bb,\cc)]-[\Delta(\bb,\cc)]\\
    &=4\Delta(\bb,\cc)/n~.
\end{align}
\end{subequations}
This quantity is bounded by $4d_X/n$, where $d_X$ is the distance of the other underlying binary linear code $\grp{C_X}$.
%\textcolor{red}{\footnotesize [should this be $C_X$?]}

\section{Performance of quKit QSCs}
\vva*{
\label{appx:performance}

\begin{figure}[h!tbp]
    \centering
    \begin{subfigure}[b]{0.45\textwidth}
        \begin{tabular}{cccc|cccc}
\toprule 
 & \multicolumn{3}{c|}{cat} & \multicolumn{4}{c}{simplex} \tabularnewline\midrule
$K$ & $d_E$ & $\nb = \nb/n$ & $F_\text{max}$ & $d_E$ & $\nb$ & $\nb/n$ & $F_\text{max}$ \tabularnewline\midrule
$2$ & $2$ & $1.6920$ & $0.9822$ & $1.5$ & $3.0776$ & $1.5388$ & $0.9841$ \tabularnewline
$3$ & $1$ & $2.6768$ & $0.9603$ & $0.8820$ & $5.3919$ & $2.6960$ & $0.9604$ \tabularnewline
$4$ & $0.5858$ & $3.6403$ & $0.9318$ & $0.8820$ & $5.3442$ & $2.6721$ & $0.9552$ \tabularnewline
$5$ & $0.3820$ & $4.5993$ & $0.8992$ & $0.6909$ & $5.1868$ & $2.5934$ & $0.9532$ \tabularnewline
$6$ & $0.2680$ & $5.5608$ & $0.8642$ & $0.4173$ & $5.5496$ & $2.7748$ & $0.9412$ 
\tabularnewline
\bottomrule
\tabularnewline
\tabularnewline

\end{tabular}
\caption{\label{tab:cat_vs_simplex_data}Sweet spot data and other code parameters are listed where \(K\) is logical dimension, $n$ is the number of modes, $\nb$ the total energy required and $F_\text{max}$ the fidelity achieved at the code's sweet spot.}

    \end{subfigure}
    \hfill
    \begin{subfigure}[b]{0.5\textwidth}
        \centering
        \includegraphics[width=\textwidth]{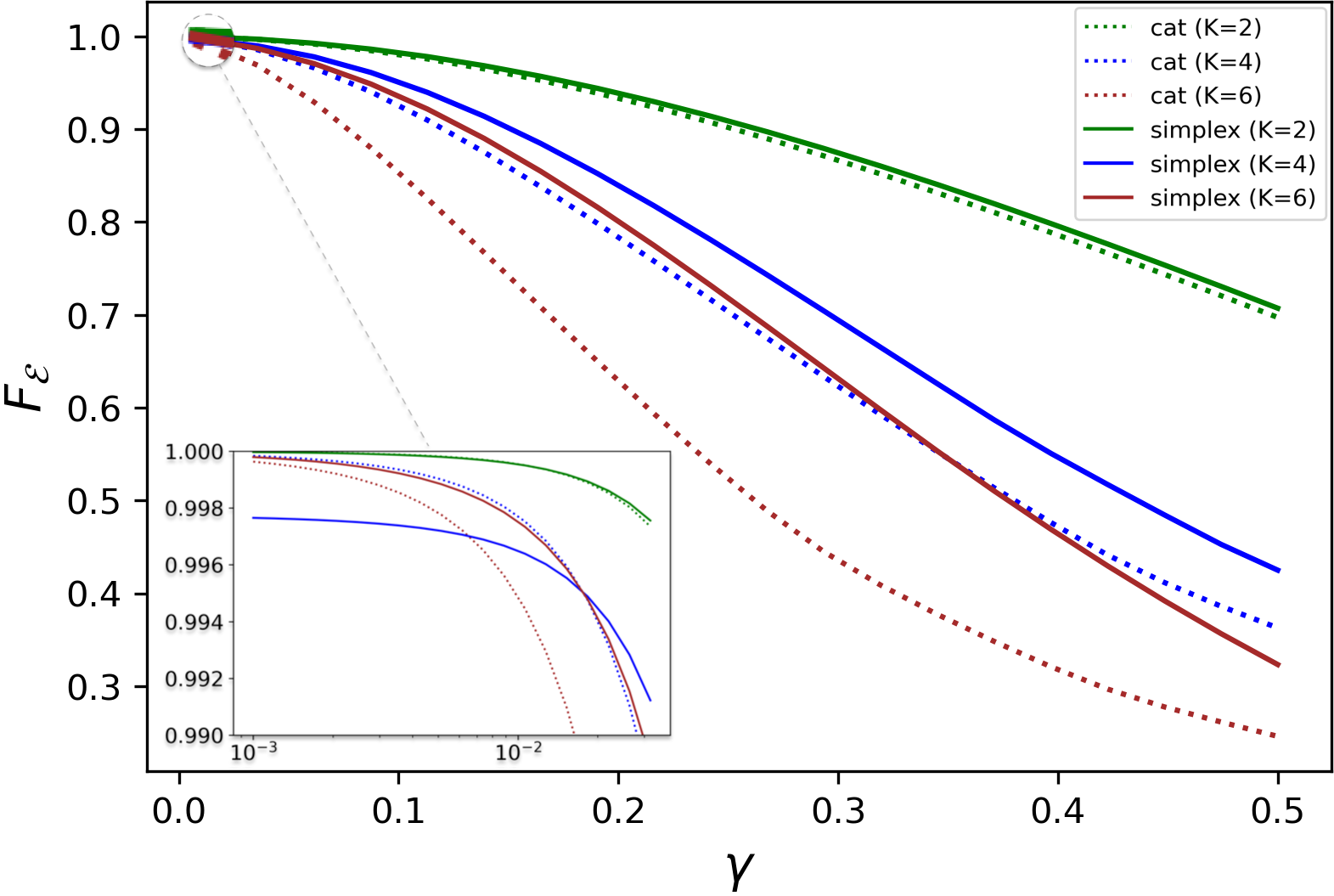}
        \caption{\label{fig:cat_vs_simplex_fids}Channel fidelity $F_\mathcal{E}$ is plotted at each code's respective sweet spot energies.} 
    \end{subfigure}
    
    \caption{Comparing cat ($1$-mode) and simplex ($2$-mode) quKit codes for varying values of $K$, it is observed that the simplex family provides more pronounced advantages in code parameters and performance with growing logical dimension. The sweet-spot energy was calculated at the loss rate $\gamma=0.095$.}
    \label{fig:cat_vs_simplex}
\end{figure}

%\color{red}
Here, we present results of a numerical comparison of several new intrinsically multi-mode poly\textit{tope} constellations to single-mode and multi-mode instantiations of cat-code (i.e., poly\textit{gon}-based) constellations.
% We discuss the advantages of employing QSCs in comparison to cat codes for encoding information in higher logical dimensions, where we observe that
% We are limited by computational power in our numerical comparisons, which do not allow us to to do computations for very large constellations.
% \spj{Higher energy computation happen faster actually. Maybe also mention number of vertices as a limitation here if we're mentioning this. That is a big limitation - perhaps more so than the dimension. } \vva{Maybe u can explain this? I'm confused why this is the case. Is this because you're using some trick associated with Anthony et al? I expressed GKP states in terms of coherent states and that had a lot of vertices, but comp time was limited to HS dimension.}\spj{Yes, because of Aurelie et al.'s trick. I'll analyse a bit more and write something up.}
We observe that multi-mode QSCs efficiently utilize the extra dimensions to store more logical information, all while consuming a comparable (in most cases, lower) energy per mode.
% \spj{In most cases, QSCs consume lower energy per mode.}
% The comparison highlights that QSCs deliver superior performance and error protection, all while consuming lower energy per mode.

% \vva{I forgot this... i think it might be more fair to also include a 3-cat times a 2-cat code? Same logical dim but still consisting of cat codes.}\spj{MK wins over this code. Products of cats in this fashion seems pretty non trivial to me though. For example, a hessian (27 pts) is the union of 3 copies of $3-\text{cat} \otimes -1*3-\text{cat}$ (9 pts). Each copy living in one of $\binom{3}{2}$ pairs of modes }

Our performance metric, as guided by \cite{schumacher_sending_1996, reimpell_iterative_2005, fletcher_optimum_2007, fletcher_channel-adapted_2008}, is the \emph{channel fidelity}
\begin{equation}
F_{\mathcal{E}} \equiv \bra\Psi|\rho_{\mathcal{E}}|\Psi\ket
\end{equation}
where, for a qubit state, $|\Psi\ket = (|0_A 0_B\ket + |1_A 1_B\ket)/\sqrt{2}$ is the maximally entangled state between the source qubit $A$ and the ancilla qubit $B$.
The outgoing density matrix,
$\rho_{\mathcal{E}} \equiv \mathcal{E}_A \otimes \mathcal{I}_B(|\Psi\ket\bra\Psi|)$,
is obtained by the action of the combined encoding-noise-recovery channel $\mathcal{E}$ on the source qubit and identity $\mathcal{I}$ on the ancilla. The channel fidelity $F_\mathcal{E}$ is an intrinsic property of the channel which measures how well the entanglement between the information qubit and an ancillary system in preserved upon application of the channel $\mathcal{E}$. For more motivation behind our choice of metric, we refer the interested reader to \cite[Appx. A]{albert_performance_2018}.

The channel $\mathcal{E}$ is considered to be the composition of the encoding, noise and recovery channels. 
We assume that noise occurs only via the pure-loss channel, described by Kraus operators \cite{nielsen_quantum_2011}
\begin{equation}
E_{\ell} \equiv\left(\frac{\gamma}{1-\gamma}\right)^{\ell / 2} \frac{\hat{a}^{\ell}}{\sqrt{\ell !}}(1-\gamma)^{\hat{n} / 2}~,
\end{equation}
where \(\ell \geq 0\) quantifies the amount of photons lost, and where $\gamma$ is the \emph{loss rate}.
For a selected encoding and this error channel, we optimize the recovery to obtain the maximum $F_\mathcal{E}$. This optimization problem can be formulated as a semidefinite program \cite{fletcher_optimum_2007}, which we solve using the Python library \href{https://www.cvxpy.org/index.html}{\texttt{CVXPY}} \cite{agrawal_rewriting_2018, diamond_cvxpy_2016}. 

The above technique can be adapted to bosonic codes by setting a maximum Fock-space cutoff (in order to make the underlying space finite-dimensional) \cite{albert_performance_2018}.
We avoid such truncation by working in the coherent-state basis.
In such a basis, the action of the pure loss channel can be expressed using a different set of Kraus operators whose cardinality and matrix dimension are equal to the size of the code constellation \cite[Appx.~A]{denys_2t-qutrit_2022}. 
That way, we are constrained more by the size of the code constellation than the number of modes.
% This allows for numerical comparisons of slightly larger multimode codes \spj{I don't think this is accurate. The trick makes the time complexity $O(\#pts^2 n^{0.73})$, roughly eyeballing some data. So we can go to higher \textbf{modes} but are limited to how large of a constellation we can consider. The wording right now suggest to me that we can go to larger number of pts than brute-force, which is sort of opposite of what the trick does.} than purely brute-force numerics.
% but computational power restricts us from studying large constellations.

A $K$-dimensional code is constructed by replicating a ``base'' logical constellation $K$ times while maintaining good resolution $d_E$. 
Cat codes use $n$-gons as the base constellations, while simplex codes employ $\mathcal{C}_0$ from Eq.~\eqref{eq:simplex}. 

The $k$th logical constellation of a $2$-gon qu\(K\)it code with \(0\leq k < K\) is generated by multiplying the base line segment $\{1,-1\}$ with $e^{\i\pi k/K}$.
The \(k\)th logical constellation \(\R_k\CC_0\) for the qu\(K\)it simplex codes with \(0\leq k < K\in\{2,3,4\}\) is obtained 
% from  the base simplex constellation \(\CC_0\) \eqref{eq:simplex} 
by letting \(\{\R_0,\R_1,\R_2,\R_3\}=\{I,-I,Z,-Z\}\), where \(I\) is the two-dimensional identity and \(Z\) is the Pauli-\(Z\) matrix.
% $\left\{\textstyle{\frac{1}{\sqrt{2}}}(\o^{\mu},\o^{2\mu})\,|\,\mu\in\grp{\bbZ_{5}}\right\}$.
The $K=5$ (\(K=6\)) simplex constellations are generated using the unitary rotations $\{\omega^k I\,|\, k \in \mathbb{Z}_5\}$ ($\{e^{2k\pi i/6} I\,|\, k \in \mathbb{Z}_6\}$).

\parg{Sweet-spot comparison}
% Logical constellations of the $2-$gon quKit codes are $e^{2k\pi i/2K}\{1,-1\},\,0\leq k \leq K$.
% For this comparison, we compare several qu\(K\)it simplex codes to 2-gon (i.e., line segment based) cat codes.

Given a loss rate \(\gamma\), one can tune the energy of a given code to obtain the \textit{sweet spot} energy value --- the \(\nb\) that gives the highest fidelity \(F_{\text{max}}\). 
For cat codes, it has been observed \cite{michael_new_2016,li_cat_2017,albert_performance_2018} that this sweet spot value is finite, and that it does not drastically change with small changes in the loss rate \(\gamma\). 
We observe similar behavior in all the QSCs we examine.

We evaluate the performance of each code at its respective sweet spot in order to compare the highest possible performance of each code under a given loss rate.
Figure~\ref{tab:cat_vs_simplex_data} lists the code parameters of the simplex and $2$-gon based cat quKit codes. 
The advantage of using simplex codes over the cat becomes pronounced for larger memories. As we scan the table in the figure, we see that the fidelity \(F_{\text{max}}\) of simplex codes decreases \textit{slower} with growing dimension \(K\) compared to that of cat codes, meaning that simplex codes utilize the available phase space more effectively when packing more quantum information. This is corroborated by the simplex qu\(K\)its maintaining higher resolution \(\de\) for large $K$. 

The energy required per mode ($\nb/n$) for optimal simplex performance also increases at a slower rate than that of cat codes. Notably, for $K=6$, even the total energy ($\nb$) needed by simplex codes is lower than that of the corresponding cat code. This trend is consistent in code performance, quantified by the channel fidelity, as shown in Fig.~\ref{fig:cat_vs_simplex_fids}.

\begin{table}[h!tbp]
    \begin{tabular}{c c c c c c c}
\toprule 
 
logical const-n & code const-n & $t_{\downarrow}$ & $d_E$ & $\nb$ & $\nb/n$ & $F_\text{max}$  \tabularnewline\midrule

$2$-gon & $12$-gon & $1$ & $0.268$ & $5.5608$ & $5.5608$ & $0.8642$  \tabularnewline
$3$-gon & $18$-gon & $2$ & $0.121$ & $8.9584$ & $8.9584$ & $0.8882$  \tabularnewline
$3\text{-gon} \otimes 3\text{-gon}$ & $9\text{-gon} \otimes 6\text{-gon}$ & $2$ & $0.234$ & $9.1801$ & $4.5901$ & $0.9585$  \tabularnewline
Möbius-Kantor & $2\{8\}3$ & $2$ & $0.367$ & $5.7992$ & $2.8996$ & $0.9901$  \tabularnewline

\bottomrule

\end{tabular}
\caption{
\label{tab:qu6it_data}Sweet spot data and other code parameters for quKit codes with $K=6$ are listed where $n$ is the number of modes, $\nb$ the total energy required and $F_\text{max}$ the fidelity achieved at the code's `sweet spot'.}
\end{table}

\begin{figure}[h!tbp]
    \centering
    \begin{subfigure}{0.45\textwidth}
        \includegraphics[width=\textwidth]{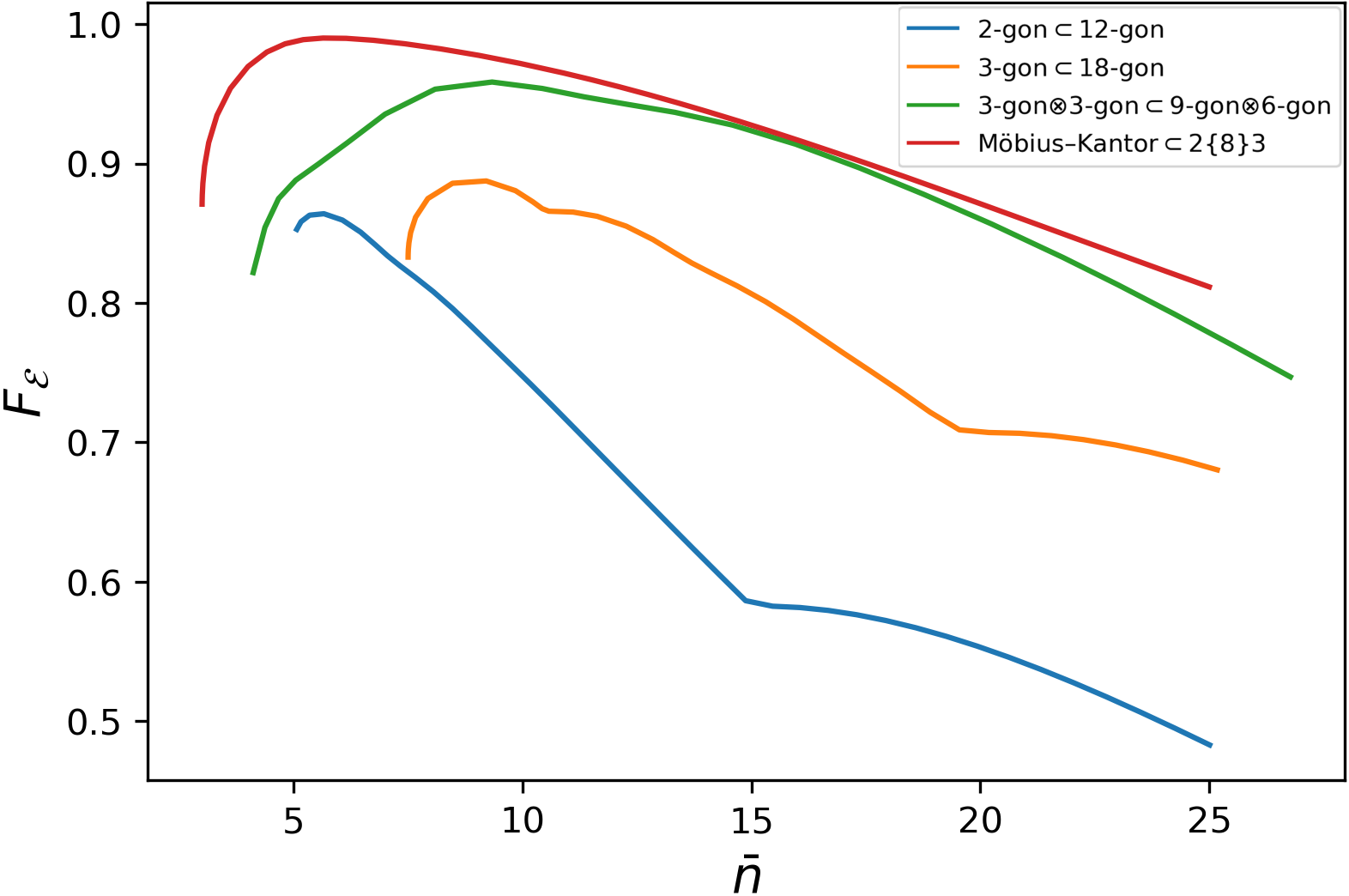}
        \caption{\label{fig:mk_universal_advantage}Channel fidelity $F_\mathcal{E}$ is plotted at fixed loss rate $\gamma=0.095$.}\
    \end{subfigure}
    \hfill
    \begin{subfigure}{0.45\textwidth}
        \includegraphics[width=\textwidth]{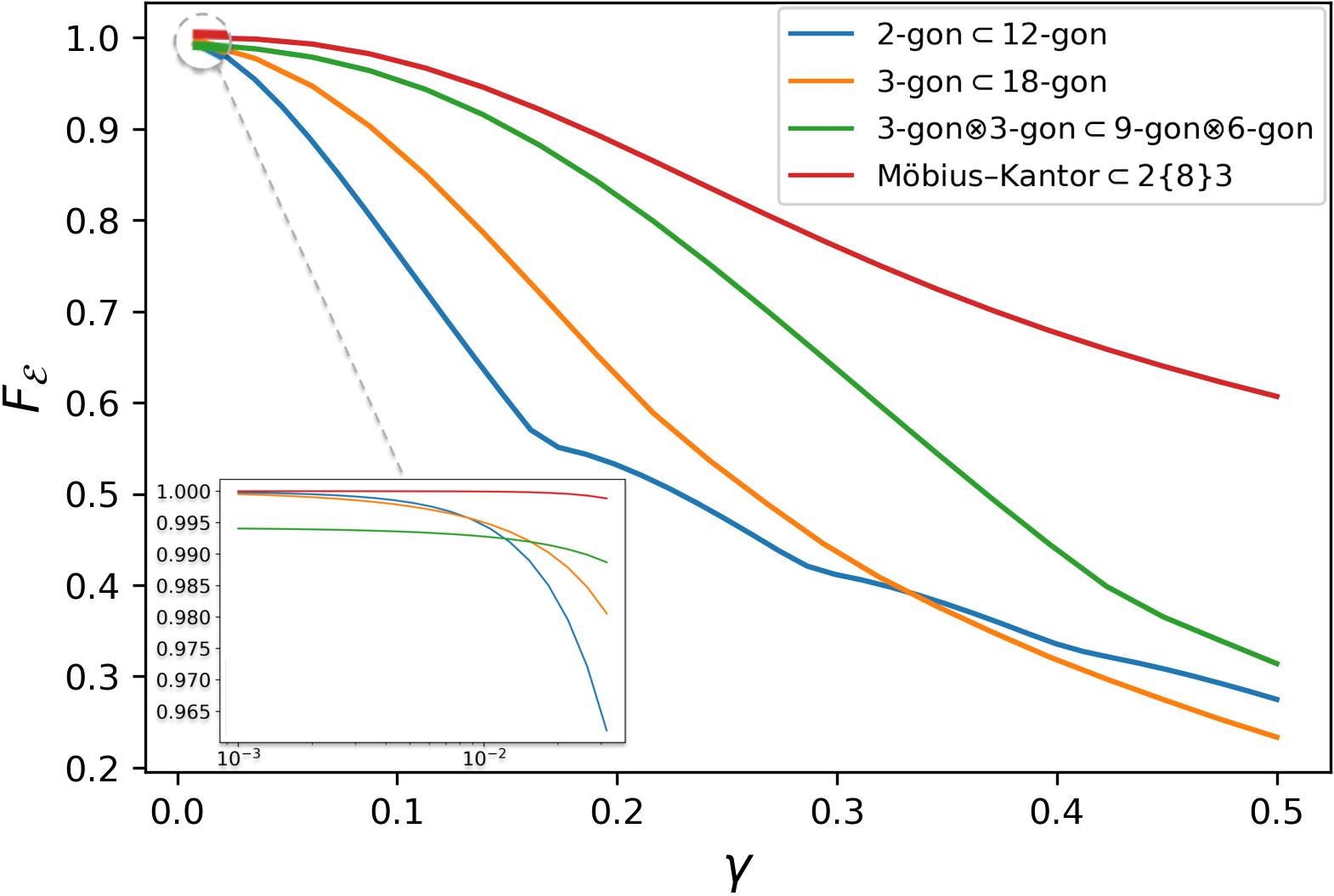}
        \caption{\label{fig:qu6it_fids_18sweetspot}Channel fidelity $F_\mathcal{E}$ is plotted at fixed total energy $\nb=8.9584$.}
    \end{subfigure}
    \caption{As shown in (a), the Möbius-Kantor code demonstrates a universal improvement over polygon based codes and outperforms them over a range of energies and loss rates, as exemplified in (b) by choosing $\nb$ corresponding to the sweet spot energy (at $\gamma=0.095$) of the $3\text{-gon}\subset 18\text{-gon}$ code.
    }
    \label{fig:codes_performance}
\end{figure}

\parg{Overall advantage of a qudit encoding}
% This advantage extends beyond simplex codes to general multimode QSCs. 
We also compare overall performance of a multi-mode QSC to various cat-like codes by sweeping both energy and loss rate.
We fix $K=6$ and construct codes out of various logical constellations: the $2$-gon, $3$-gon ($\{1,e^{2\pi i/3},e^{4\pi i/3}\}$), $3\text{-gon}\otimes 3\text{-gon}$($\{(e^{2\pi im_1/3},e^{2\pi im_2/3})\,|\,0\leq m_1,m_2\leq 2\}$) and the Möbius-Kantor polygon.
The first two are single-mode cat codes, the third distributes logical information over two modes using tensor products of single-mode cat codes, while the last is an intrinsically two-mode code.

Results from a numerical comparison in Fig.~\ref{fig:mk_universal_advantage} show a universal advantage across the swept energy-and-loss-rate parameter space.
% , showing that the Möbius-Kantor code excels across a wide range of $\nb$ values at a fixed loss rate $\gamma$. 
Similar trends are observed for various other $\gamma$ values (not shown here). Notably, the Möbius-Kantor surpasses other codes even at their optimal values, as exemplified in Fig.~\ref{fig:qu6it_fids_18sweetspot}, where we choose the $\nb$ corresponding to the $3$-gon $\subset$ $18$-gon code's sweet spot.

We append sweet-spot data for this set of codes in Tab.~\ref{tab:qu6it_data}, which corroborates the simplex-cat-code data in Fig.~\ref{tab:cat_vs_simplex_data}.
We observe that the Möbius-Kantor code provides robust protection against up to 2 losses, boasts higher resolution ($d_E$), requires lower energy per mode, and consistently outperforms all the mentioned polygon based codes in the sweet-spot comparison.

When encoding a greater number of logical dimensions, multimodal QSCs prove significantly more resource efficient and clearly outperform cat codes.
% , which are limited to single-mode utilization.

\color{black}

% %%%%%%%%%%%%%%%%%%

}
\section{Lindbladian stabilization}
\vva*{
\label{appx:lindblad}

The $Z$-type (i.e., dephasing or rotation error) correction for cat codes is done autonomously by engineering
Lindbladians with a desired ``correcting'' jump operator $F=\kappa(a^{2p}-\nb^{p})$ and correction rate \(\kappa\).
Engineering such terms is possible in microwave cavities coupled to superconducting circuits \cite{girvin_introduction_2023}.
The typical scheme proceeds by coupling the physical system to an ancillary
or buffer mode $b$ via the Hamiltonian coupling $Fb^{\dg}+\text{H.c.}$,
setting the ancillary mode to have a high loss rate, and then showing that the effective
Lindbladian acting on the physical \(a\)-mode system has jump operator
$F$ (see, e.g., Ref.~\cite{mirrahimi_dynamically_2014}).

Multi-mode coherent-state QSCs require more jump operators, and each jump
operator can now consist of multiple monomials in the lowering operators
$a_{j}$. 
However, due to the flexibility provided by a recently
developed circuit element called an \textit{asymmetrically threaded SQUID}, or \textit{ATS} \cite{lescanne_exponential_2020}, the above scheme
can be extended to realize the more complex jumps required for QSCs. 
We sketch out a general scheme below and apply it to a CSS and a polytope QSC. 

The cost of our basic scheme --- one ancillary mode per jump ---
is only an upper bound. While more advanced schemes are outside
the scope of this work, we note that a single ATS can be used to simulataneously realize \textit{multiple} jumps using as little as one ancillary mode \cite[Appx.~B.2]{chamberland_building_2022}.

\subsubsection{General scheme}

A desired jump operator is a sum of monomials of some maximum degree
and a potential constant term that is the $P$th power of $\nb$,
\begin{equation}
F=\sum\text{monomial(\ensuremath{a_{j}})}-\text{constant}\cdot\nb^{P}~.\label{eq:jump-general}
\end{equation}
Leveraging previous schemes \cite[supplement]{lescanne_exponential_2020}\cite[Appx.~B.2]{chamberland_building_2022}, we describe a slightly more general scheme to implement a dissipator with this
jump operator, which generates time evolution according to the equation of motion (\ref{eq:dissipator}).
%providing examples with specific parameters afterwards.

Let the harmonic component of the $j$th physical mode have frequency
$\omega_{j}$, while the ancilla \(b\)-mode evolves at $\omega_{b}$.
In the rotating frame w.r.t. these components, the multi-mode
density matrix $\rho_{ab}$ describing a set of modes coupled via an ATS evolves according to
\begin{equation}
\dot{\rho}_{ab}=-i[H_{\text{drive}}+H_{\text{ATS}},\rho]+\kappa_{b}{\cal D}[b](\rho)\,.
\end{equation}
We describe each term and its purpose:
\begin{enumerate}
\item The drive term, $H_{\text{drive}}=-\nb^{P}b+\text{H.c.}$, will yield
the constant part of the jump operator $F$ (\ref{eq:jump-general})
once the effective equation of motion on the physical modes is derived.
This term can be set to zero if no constant term is necessary.
\item All of the magic comes from the ATS term \cite[Eq.~(B11)]{chamberland_building_2022}\footnote{Up to constant prefactors, this term is also obtained from the last two terms in \cite[Eq.~(S2)]{lescanne_exponential_2020} by setting \(\varphi_\Delta = \varphi + \pi/2\), \(\Delta E_J =0\), and \(\varphi_\Sigma = \epsilon(t) + \pi/2\), and expanding to first order in \(\epsilon\). 
The quadratic term in that equation can absorbed into the bare oscillator Hamiltonian prior going into the rotating frame by expanding \(\varphi\) in a set of normal modes.},
\begin{equation}\label{eq:EOM}
H_{\text{ATS}}=\epsilon(t)\sin\left(\varphi+\dr_{b}be^{-i\omega_{b}t}+\mathord{\sum}_{j}\dr_{j}a_{j}e^{-i\o_{j}t}+\text{H.c.}\right)\quad\text{with pump tones}\quad\epsilon(t)=\mathord{\sum}_{p}\tone_{p}e^{i\Omega_{p}t}+\text{H.c.}\,,
\end{equation}
which depends on static real parameters $\{\dr_{b},\dr_{j},\varphi\}$ and tunable real parameters $\{\omega_{b},\omega_{j},\tone_{p},\Omega_{p}\}$.
The static flux $\varphi\in\{0,\pi/2\}$ \cite[Eq.~(S3)]{lescanne_exponential_2020} allows us to interpolate between a sine
and cosine ATS term.
% The number of physical mode frequencies $\omega_{j}$ present in the interaction depends on the support of the jump. 
One
pump tone, with amplitude \(\tone_p\) and frequency $\Omega_{p}$, is necessary for each
monomial in the jump operator (\ref{eq:jump-general}). Tuning the frequency allows us to select the specific desired monomial, while tuning the drive allows us to tune the monomial's coefficient.
\item The dissipative part, $\kappa_b{\cal D}[b]$ for sufficiently large
$\kappa_b>0$, ensures that the ancilla is sufficiently lossy. The steady-state space of this evolution is spanned by any state of the \(a_j\) modes, tensored with the vacuum Fock state \(|0\rangle\) on the \(b\) mode.
Assuming the Hamiltonian terms to be perturbations to this strong Lindbladian evolution, one can then derive an effective equation of motion within this steady-state space using either second-order perturbation theory or what is colloquially known as ``adiabatic elimination'' \cite{reiter_effective_2012,azouit_adiabatic_2016,azouit_towards_2017}.
\end{enumerate}
Expanding the ATS term yields an infinite series, with combinatorially
many monomials consisting of products of drive-tone terms $\{\tone_{p}e^{i\Omega_{p}t}\}$,
physical mode operators $\{a_{j}e^{-i\o_{j}t}\}$, the ancillary mode term
$b e^{-i\omega_{b}t}$, and the flux bias $\varphi$. The expansion is approximated
by truncating to an order such that the highest-degree term is one
higher than the degree of the highest-order monomial in the desired jump operator \eqref{eq:jump-general}.
The phase term $\varphi\in\{0,\pi/2\}$ ensures that the expansion
contains the monomial of correct (even or odd) degree. 

The pump-tone frequencies $\{\Omega_{p}\}$ are then tuned to particular
linear combinations of $\{\omega_{j},\omega_{b}\}$ so that any of the
monomials that are also present in the desired jump operator become time-\textit{in}dependent.
That way, all other terms can be treated as higher-order ``fast-rotating''
corrections in what is known as the ``rotating-wave approximation''.
Combining with the drive term, the Hamiltonian terms in Eq.~\eqref{eq:EOM} are then approximated by \(Fb^\dagger+\text{H.c.}\).
Verifying that the many remaining terms in the expansion are all time-\textit{de}pendent can be done using
the algebraic manipulation plugin \textsc{sneg} \cite{zitko_sneg_2011,zitko_rokzitkosneg_2022} in \textsc{Mathematica}.

The desired equation of the density matrix $\rho=\text{tr}_{\text{mode }b}(\rho_{ab})$
on the physical modes upon adiabatically eliminating the ancilla is
then
\begin{equation}
\dot{\rho}=\kappa{\cal D}[F](\rho)+\cdots\,,\label{eq:dissipator}
\end{equation}
for a to-be-determined correction rate $\kappa$, and up to higher-order
corrections ``$\cdots$'' stemming from corrections to the approximations.

\subsubsection{CSS QSCs}

Our $Z$-type stabilization for CSS-type concatenated encodings
provides an autonomous alternative to the discrete measurement of
$Z$-type error syndromes. Such jump operators can be readily ``plugged
in'' to any concatenated cat-CSS code, including a recent concatenated
surface-cat code proposal \cite{chamberland_building_2022}.

Our jump operator for the surface-cat code example consists of a product
of lowering operators acting on sides one through four of each plaquette, $F=a_{1}a_{2}a_{3}a_{4}-\nb^{2}$, of a square lattice.
This is a special case of the general form (\ref{eq:jump-general})
with one degree-four monomial and constant term with $P=2$.
The sole monomial requires only one pump tone, with amplitude \(\tone_1\equiv \tone\) and frequency $\Omega_{1}\equiv\Omega$, and zero
flux bias, $\varphi=0$. The ATS sine term is expanded to fifth
order. The condition selecting the desired monomial is
\begin{equation}
\Omega_{1}=\omega_{1}+\omega_{2}+\omega_{3}+\omega_{4}-\omega_{b}\,,\quad\quad\text{yielding the monomial}\quad\quad a_{1}a_{2}a_{3}a_{4}b^{\dg}+\text{H.c.}\,.
\end{equation}
This monomial is multiplied by a product of accompanying constants, $\dr_{1}\dr_{2}\dr_{3}\dr_{4}\dr_b \tone$,
to yield an effective correction rate $\kappa\propto(\dr_{1}\dr_{2}\dr_{3}\dr_{4}\dr_b \tone)^{2}/\kappa_{b}$,
after adiabatic elimination.

The above scheme is done for each plaquette of the surface-code architecture.
The additional $a_{j}^{2}-\nb$ dissipators --- required for restricting each mode \(j\) to antipodal coherent states --- 
are realized in said architecture using a single ATS \cite[Appx.~B.2]{chamberland_building_2022}. 
Together, these provide autonomous protection
against \textit{all} $Z$-type errors, utilizing the full error-correcting
power of the (outer) surface code for such noise.
Extension to other QLDPC codes is straightforward, barring any issues with long-range physical connectivity.

\subsubsection{Hessian QSC}

The Hessian code requires three jump operators, two of which consist of three monomials.
Each jump operator can be realized using the general scheme above,
providing a non-trivial QSC example that should be realizable with state-of-the-art
ATS technology.
\begin{enumerate}
\item The jump $F=a_{1}a_{2}a_{3}$ has no constant term, so no drive term
is necessary. Only one pump tone, with amplitude \(\tone_1\equiv \tone\) and frequency $\Omega_{1}\equiv\Omega$, is required,
and $\varphi=\pi/2$ to obtain a \textit{cosine} ATS.
The ATS term is expanded to second order. The conditions selecting the
desired monomial are
\begin{equation}
\Omega=\omega_{1}+\omega_{2}+\omega_{3}-\omega_{b}\quad\quad\text{yielding the monomial}\quad\quad a_{1}a_{2}a_{3}b^{\dg}+\text{H.c.}\,.
\end{equation}
This monomial corresponds to an effective correction rate $\kappa\propto(\dr_{1}\dr_{2}\dr_{3}\dr_b \tone)^{2}/\kappa_{b}$
after adiabatic elimination.
\item The jump $F=a_{1}^{3}+a_{2}^{3}+a_{3}^{3}$ also has no constant term.
The three terms require three drive tones, with parameters \(\{\tone_{p},\Omega_{p}\}\) for \(p\in\{1,2,3\}\).
The phase $\varphi=\pi/2$ so that the ATS term becomes a cosine.
The ATS cosine term is expanded to second order. The conditions selecting
the desired monomials are
\begin{equation}
\begin{aligned}\tone_{p} & =1/\dr_{j=p}^{3}\\
\Omega_{p} & =3\omega_{j=p}-\omega_{b}
\end{aligned}
,\quad\quad\text{yielding the monomials}\quad\quad a_{j=p}^{3}b^{\dg}+\text{H.c.}\,.
\end{equation}
Each of these monomials is multiplied by a product of respective accompanying constants,
$\dr_{j=p}^{3}\dr_b\tone_p = \dr_b$, where we have used the drive-tone amplitudes to cancel the non-tunable coupling strengths \(\dr_j\).
This yields an effective correction rate $\kappa\propto\dr_b^{2}/\kappa_{b}$,
after adiabatic elimination.
\item The jump $F=a_{1}^{6}+a_{2}^{6}+a_{3}^{6}-\nb^{3}/4$ has three monomials
and a constant term with power \(P=3\) and coefficient \(1/4\). 
The three monomials require three drive tones,
with parameters \(\{\tone_{p},\Omega_{p}\}\) for \(p\in\{1,2,3\}\). The phase $\varphi=0$ so that
the ATS term remains a sine. This term is then expanded to fourth
order. The conditions selecting the desired monomials are
\begin{equation}
\begin{aligned}\tone_{p} & =1/\dr_{j=p}^{6}\\
\Omega_{p} & =6\omega_{j=p}-\omega_{b}
\end{aligned}
,\quad\quad\text{yielding the monomials}\quad\quad a_{j=p}^{6}b^{\dg}+\text{H.c.}\,.
\end{equation}
Each of these monomials is multiplied by $\dr_b $.
The fluxes \(g_j\) are required to be equal for all three \(j\) in order to realize the jump.
This yields an effective correction rate $\kappa\propto\dr_b^{2}/\kappa_{b}$,
after adiabatic elimination.
\end{enumerate}

}

%\pagebreak
\twocolumngrid
\bibliography{references}

\end{document}